\documentclass[fleqn,usenatbib]{mnras}
\usepackage{newtxtext,newtxmath}
\usepackage[T1]{fontenc}

\DeclareRobustCommand{\VAN}[3]{#2}
\let\VANthebibliography\thebibliography
\def\thebibliography{\DeclareRobustCommand{\VAN}[3]{##3}\VANthebibliography}
\usepackage{graphicx}	
\usepackage{amsmath}	

\newcommand{\equ}[1]{eq.~(\ref{eq:#1})}
\newcommand{\se}[1]{\S\ref{sec:#1}}
\newcommand{\fig}[1]{Fig.~\ref{fig:#1}}
\newcommand{\Fig}[1]{Figure~\ref{fig:#1}}
\newcommand{\be}{\begin{equation}}
\newcommand{\ee}{\end{equation}}

\newcommand{\ssim}{\!\sim\!}
\newcommand{\seq}{\!=\!}
\newcommand{\sgt}{\!>\!}
\newcommand{\slt}{\!<\!}
\newcommand{\sdash}{\!-\!}

\newcommand{\msun}{\,\rm M_\odot}
\newcommand{\kms}{\ifmmode\,{\rm km}\,{\rm s}^{-1}\else km$\,$s$^{-1}$\fi}
\newcommand{\Mpc}{\,{\rm Mpc}}
\newcommand{\kpc}{\,{\rm kpc}}
\newcommand{\pc}{\,{\rm pc}}
\newcommand{\Gyr}{\,{\rm Gyr}}
\newcommand{\Myr}{\,{\rm Myr}}
\newcommand{\yr}{\,{\rm yr}}
\newcommand{\ergs}{\,{\rm erg}\,{\rm s}^{-1}}

\newcommand{\Mv}{M_{\rm v}}
\newcommand{\Rv}{R_{\rm v}}
\newcommand{\Vv}{V_{\rm v}}
\newcommand{\Mg}{M_{\rm g}}
\newcommand{\Ms}{M_{\rm s}}
\newcommand{\Vsn}{V_{\rm SN}}
\newcommand{\tdep}{t_{\rm dep}}
\newcommand{\tinf}{t_{\rm inf}}
\newcommand{\Mbh}{M_{\rm bh}}
\newcommand{\dotMbh}{\dot{M}_{\rm bh}}
\newcommand{\epsr}{\epsilon_{\rm r}}

\title[Compaction-Driven Black Hole Growth]
{Compaction-Driven Black Hole Growth}

\author[Lapiner, Dekel, Dubois]
{Sharon Lapiner$^1$\thanks{E-mail: sharon.lapiner@mail.huji.ac.il},
Avishai Dekel$^{1,2,3}$,
Yohan Dubois$^{3}$
\\
$^1$Racah Institute of Physics, The Hebrew University, Jerusalem 91904 Israel\\
$^2$SCIPP, University of California, Santa Cruz, CA 95064, USA\\
$^3$Institut d'Astrophysique, 98 bis Boulevard Arago, 75014 Paris, France\\ 
}

\date{Accepted 2021 April 23. Received 2021 April 23; in original form 2020 December 10}
\pubyear{2021}

\begin{document}

\label{firstpage}
\pagerange{\pageref{firstpage}--\pageref{lastpage}}
\maketitle
\large
\begin{abstract}
We study the interplay between galaxy evolution and central black-hole (BH) 
growth using the {NewHorizon} cosmological simulation. 
BH growth is slow when the dark-matter halo is below a golden mass of 
$\Mv \ssim 10^{12}\msun$, and rapid above it. The early suppression is 
primarily due to gas removal by supernova (SN) feedback in the shallow 
potential well, predicting that BHs of $\sim\! 10^5\msun$ tend to lie below the 
linear relation with bulge mass.
Rapid BH growth is allowed when the halo is massive enough to lock in the SN 
ejecta by its deep potential well and its heated circum-galactic medium (CGM).
The onset of BH growth between these two zones is triggered by a 
wet-compaction event, caused, e.g., by mergers or counter-rotating streams.  
It brings gas that lost angular momentum into the inner-$1\kpc$ ``blue nugget" 
and causes major transitions in the galaxy structural, kinematic and 
compositional properties, including the onset of star-formation quenching.  
The compaction events are confined to the golden mass by the same mechanisms of
SN feedback and hot CGM.  The onset of BH growth is associated with its sinkage
to the center due to the compaction-driven deepening of the potential well and 
the associated dynamical friction. 
The galaxy golden mass is thus imprinted as a threshold for rapid BH growth, 
allowing the AGN feedback to keep the CGM hot and maintain long-term quenching.
AGN feedback is not causing the onset of quenching; they are both caused by a 
compaction event when the mass is between the SN and hot-CGM zones.
\end{abstract}
\section{Introduction}
\label{sec:intro}

Observations of central black holes (BHs) and the associated active galactic
nuclei (AGN), as well as cosmological simulations, indicate a characteristic 
mass for black holes and their host galaxies, near a stellar mass of 
$\Ms \ssim 10^{10}\msun$ or a halo mass $\Mv\ssim 10^{12}\msun$.  
For example, AGN are observed to dominate the emission lines in BPT diagrams 
\citep{baldwin81} in galaxies above this characteristic mass
\citep{kauffmann03_agn,vitale13}.
A comparable mass is indicated in studies of AGN and their host galaxies
\citep[e.g.][]{juneau15,kocevski17}.
Most notably,
a similar mass threshold is obtained for AGN-driven outflows \citep{forster19}.
At low masses, there seem to be partial indications that the masses of 
black holes 
below $10^6\msun$ fall short of the standard linear relation between black-hole 
mass and its host bulge mass \citep{kormendy13,reines15}. 
All these indicate that black-hole growth is slow in the low-mass zone
and it becomes rapid in the high-mass zone. 
This in turn indicates that AGN feedback is less relevant in low-mass galaxies
while it may be involved in the quenching of star formation in
high-mass galaxies.
Since we are not aware of an obvious feature in black-hole physics that hints 
to the origin of such a characteristic mass in the evolution of black holes,
one may suspect that this mass arises from the interplay between the black-hole
and other physical processes in the host galaxy that control the gas supply 
to the black-hole.
A better understanding of the origin of this interplay between galaxy and 
black-hole evolution is a primary issue in galaxy formation and our goal here.

\smallskip 
Indeed, galaxy formation has a golden mass at the indicated scale, which is
seen observationally and fairly understood theoretically 
\citep[e.g.][]{dekel19_gold}.
The efficiency of galaxy formation within dark-matter (DM) haloes,
as interpreted from the stellar-to-halo mass ratio ($\Ms/\Mv$)
that is derived using
abundance matching of observed galaxies and theoretical $\Lambda$CDM haloes,
shows a pronounced peak near the golden mass, with only little redshift
dependence at least in the redshift range $z\seq 0\sdash 4$  
\citep{moster10,moster13,behroozi13,rodriguez17,moster18,behroozi19}.
The decline of galaxy-formation efficiency on both sides of the peak
toward lower and higher masses indicate the operation of effective 
mechanisms that suppress star formation in the two zones, below and above the 
golden mass.
These quenching mechanisms are assumed to be supernova feedback 
(and other stellar feedback) in the low-mass zone 
\citep[e.g.][]{larson76,ds86}, 
and virial shock heating of the CGM \citep[e.g.][]{ro77,silk77,binney77,db06}
plus AGN feedback \citep[e.g.][]{croton06,cattaneo07,dubois11} 
in the high-mass zone.

\smallskip 
The same golden mass is imprinted as a golden time
in the observed cosmological evolution of star-formation-rate (SFR) density.
It is rising from $z\ssim 10$ to a broad peak near $z\ssim 2$ 
and is steeply declining after $z \ssim 1$ till the present \citep{madau14}.
The SFR-peak time at $z\ssim 1\sdash2$ reflects the same golden mass scale 
that is indicated by the stellar-to-halo ratio \citep{dekel19_gold}.
This is because, as obtained from the Press-Schechter formalism and confirmed
in cosmological simulations, the steeply rising typical halo mass that forms 
at a given time reaches a value comparable to the golden mass at 
$z\ssim 1\sdash 2$.
While the density of accretion rate into haloes is steeply declining with time 
at all epochs, reflecting the expansion of the Universe and the evolution of
the halo population \citep{dekel13},
the rise in time of the SFR density prior to its peak at $z\ssim 2$
must be obtained by quenching of galaxy formation in the low-mass haloes that 
dominate at those high redshifts \citep{bouche10}.
The drop in SFR density with time at low redshifts
is steeper than the decline associated with the accretion rate due
to quenching in the high-mass haloes that dominate at that epoch.
Thus, the rise and fall of the SFR density with time is largely determined
by the same quenching mechanisms that operate at low and high mass scales
to generate the peak in $\Ms/\Mv$.
The peak epoch of SFR density can be interpreted as a manifestation
of the golden mass scale for peak efficiency of galaxy formation. 

\smallskip 
The golden mass marks a general bimodality about $\Ms \ssim 10^{10.5}\msun$
in many galaxy properties including the characteristics of star formation, 
morphology, kinematics and composition \citep[e.g.][]{db06}.
Most relevant for the black-hole growth is a dramatic sequence of events that
typically occurs in galaxies when they are near the golden mass,
seen both in simulations
\citep{zolotov15,tacchella16_prof,tacchella16_ms,tomassetti16}
and observations \citep{barro13,dokkum15,barro17,huertas18}.
Many galaxies undergo significant gaseous compaction events
into compact star-forming ``blue nuggets".
The major compaction events that occur near the golden mass
trigger inside-out quenching of star formation,
which is maintained by the hot CGM and AGN feedback, 
leading to today's early-type galaxies.
The blue-nugget phase is responsible for dramatic transitions in the main galaxy
structural, kinematic and compositional properties,
such as a transition from dark-matter to baryon dominance in
the galaxy central $1\kpc$.
 
\smallskip 
The picture addressed here is that supernova feedback and the hot CGM,
which define the zones of low-mass and high-mass quenching of star formation,
also define the zones of low-rate and high-rate of black-hole growth
\citep{bower17,dekel19_gold}.
We propose that the same processes confine the major compaction events to 
the golden mass between these zones, and emphasize in particular that the 
compaction process is the actual trigger for a boost in the black-hole 
growth once it is in the zone where growth is possible.
The rapidly growing black-hole, through AGN feedback, can then help the hot CGM 
maintain the quenching of star formation above the golden mass.

\smallskip
The outline of this paper is as follows.
In \se{golden_compaction} we summarize the roles of supernova feedback and 
hot CGM 
in determining the golden mass and the events of wet compaction to blue
nuggets near this golden mass.
In \se{bh} we address the correlation between the compaction events and 
black-hole growth in the {NewHorizon} simulation, which is described in 
\se{app_NH}.
In \se{quenching} we discuss the implications on the quenching of star
formation.
In \se{conc} we summarize our conclusions.

\section{The golden mass and compaction to blue nuggets}
\label{sec:golden_compaction}

\subsection{Supernova feedback and virial shock heating}
\label{sec:sn}

An upper limit for a DM-halo virial velocity (which can be translated to halo
mass), within which supernova feedback can be effective in heating or ejecting 
the gas and thus suppressing the SFR, can be estimated in a simple way
using the standard theory for supernova bubbles \citep{larson76,ds86}.
The energy deposited in the ISM by supernovae that arise from a stellar mass
$\Ms$ is estimated to be
$E_{\rm SN} \ssim \Ms \Vsn^2$ with $\Vsn \ssim 120 \kms$.
The derivation involves the ratio of two timescales which turns out to be
roughly constant,
the duration of the adiabatic phase of the supernova bubble
in which it can deposit energy in the ISM before it cools radiatively,
and the dynamical timescale associated with star formation.

\smallskip
For the supernova energy to heat or eject most of the gas of mass $\Mg$
that has accreted into the galaxy, it should be comparable to the binding
energy of this gas in the DM-halo potential well,
$E_{\rm CGM} \ssim \Mg \Vv^2$, with $\Vv^2 = {G\Mv}/{\Rv}$, 
where $\Vv$ is the halo virial velocity and $\Mv$ and $\Rv$ are 
its virial mass radius.
At the peak of star-formation efficiency, if a large fraction of the gas
that has been accreted into the halo turned into stars with little ejection, 
one has $\Ms \ssim \Mg$.
By comparing $E_{\rm CGM}$ and $E_{\rm SN}$,
this yields a critical upper limit for the virial velocity of a halo
in which supernova feedback is effective,
\be
\Vv \sim \Vsn \sim 120 \kms \, .
\ee
A similar estimate arises from the velocities of supernovae in star-forming 
clumps, with an assumed SFR efficiency of a few percent and a supernova energy
efficiency of $10\!-\!20$ per cent \citep{dubois15}.
Using the standard virial relation, $\Vsn$
corresponds to $\Mv \sim 10^{11.7}\msun$ at $z=0$,
and it roughly scales as $(1+z)^{-3/2}$ in the Einstein-deSitter regime
(roughly valid at $z\sgt 1$).
In haloes above this mass, the potential well is too deep for the
supernova-driven winds to significantly heat the gas or escape.
This mass scale roughly coincides with the observed peak of star-formation
efficiency at $z \ssim 0\sdash 2$.

\begin{figure} 
\centering
\includegraphics[width=0.477\textwidth]{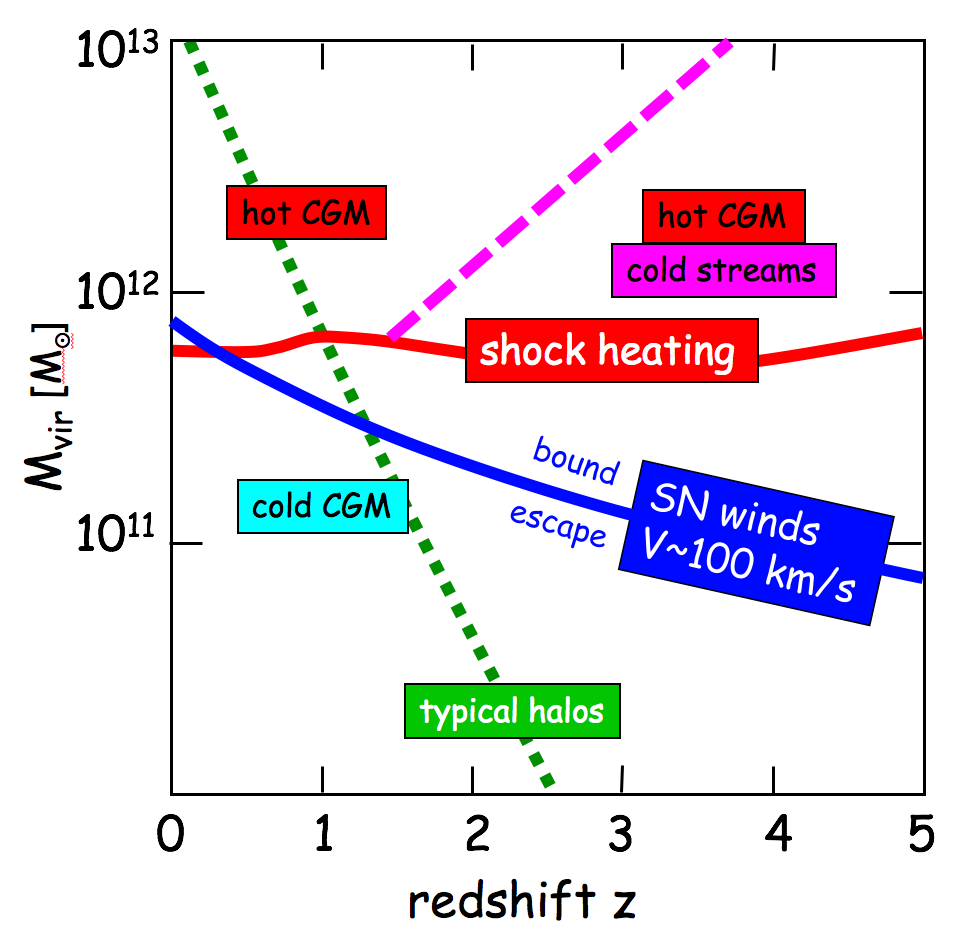}
\caption{
The golden mass of galaxy formation. 
Shown are crude estimates for the different characteristic halo masses as 
a function of redshift.
The blue curve marks the upper limit for effective supernova feedback
at $\Vv\ssim 100\kms$ based on \citet{ds86}.
The red curve shows the threshold for virial shock heating of the CGM
slightly below $10^{12}\msun$ based on \citet{db06},
with the dashed magenta line referring to the upper limit for
penetrating cold streams through the hot CGM.
The green curve refers to the Press-Schechter mass, the rapidly evolving
typical mass of forming haloes.
The left portion of the red curve at $z \slt 2$, 
combined with the magenta curve at $z \sgt 2$, 
mark the effective upper limit for star-forming galaxies.
The three characteristic masses roughly coincide at $z \ssim 1 \sdash 2$, 
defining the golden mass and the peak epoch of star formation.
}
\label{fig:scale_z}
\end{figure}

\smallskip 
\Fig{scale_z} shows the critical halo mass corresponding to $\Vv = \Vsn$ 
as a function of redshift, rising from $\sim\!10^{11}\msun$ at $z\seq 5$ 
to $\sim\!10^{12}\msun$ at $z\seq 0$.
It coincides with the more steeply rising Press-Schechter mass that 
characterizes the typical mass for haloes at $z \ssim 1\sdash 2$.

\smallskip
The other characteristic scale is the upper limit for the halo mass within
which efficient cold inflow can supply gas for star formation 
\citep{ro77,silk77,binney77}.
It is obtained by comparing the gas radiative cooling time to the relevant
dynamical time for gas inflow.
The key question addressed by \citet{bd03} and \citet{db06}
is whether the shock that forms at the halo virial radius,
behind which the gas heats to the virial temperature,
can be supported against gravitational collapse.
For the post-shock gas to be able to sustain the pressure that supports
the shock against gravity, its cooling time has to be longer than the
dynamical time for gas compression behind the shock, which is comparable to the
halo crossing time, $\Rv/\Vv$.
Since the cooling time is an increasing function of halo mass,
the shock-stability analysis reveals a mass threshold for a hot CGM
on the order of
\be
\Mv \sim 10^{11.7}\msun \, .
\ee
Its value is roughly independent of redshift in the range $z\seq 0\sdash 3$,
with an uncertainty of a factor of a few due to the uncertainty in metallicity
and the location within the halo where the shock stability is evaluated.
This analysis has been supported by idealized spherical simulations.

\smallskip 
\Fig{scale_z} also shows the predicted critical halo mass for virial shock
heating as a function of redshift, based on \citet[][Fig. 7]{db06}.
Below the critical curve one expects the cosmological inflow to be all cold,
at $T \ssim 10^4$K, efficiently feeding the galaxy and allowing high SFR.
Above the curve one expects the CGM to be shock heated to the virial
temperature, thus suppressing the cold gas supply into the galaxy, and
maintaining long-term quenching \citep{bdn07}.

\smallskip
At $z\! \geq\! 2$, above the shock-heating curve and below the dashed curve,
narrow cold streams are expected to
penetrate through the otherwise hot CGM and supply gas for efficient star
formation even in haloes above the critical mass \citep{db06}.
These predictions,
based on an analytic study of virial shock stability,
have been confirmed in cosmological
simulations \citep[][]{keres05,ocvirk08,dekel09,nelson13,nelson16}.

\smallskip 
The predictions for an upper-limit mass for cold gas supply,
as summarized in \fig{scale_z}, including the cold streams in a hot CGM
at high redshifts, are consistent with observations.
For example, 
the observed transition from star-forming galaxies to passive galaxies
in the plane of halo mass versus redshift,
based on abundance matching of galaxies to DM haloes
in a $\Lambda$CDM cosmology \citep[e.g.][Figure 13]{behroozi19}.
The possible involvement of AGN feedback in the quenching of massive galaxies
will be discussed in \se{quenching}.

\subsection{Wet Compaction to a Blue Nugget}
\label{sec:compaction}

\begin{figure*} 
\centering
\includegraphics[width=0.51\textwidth,height=0.44\textwidth,trim={1cm 0.69cm 0.cm 0cm}
]
{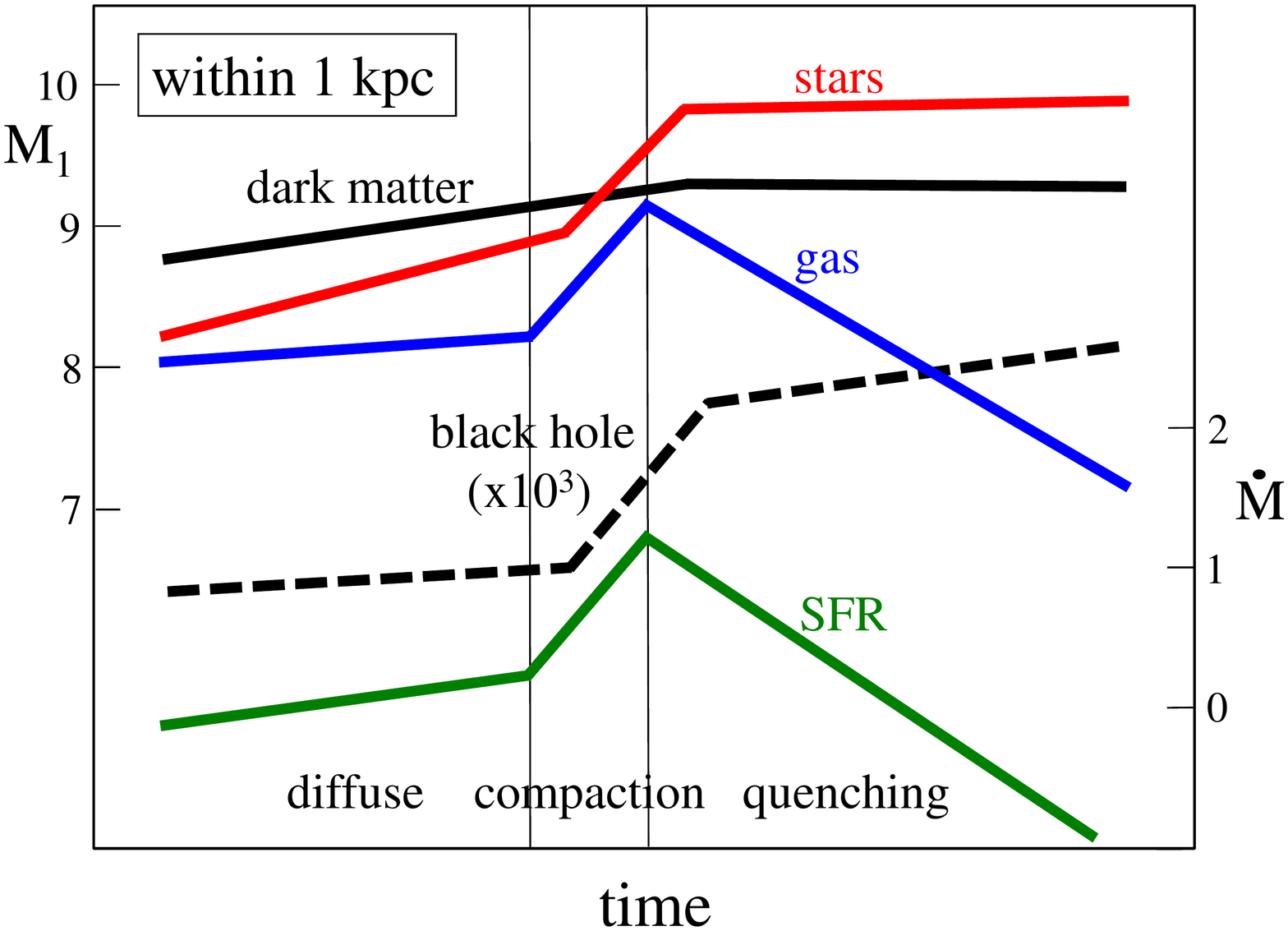}
\includegraphics[width=0.463\textwidth,trim={0 0.3cm 0 0}]
{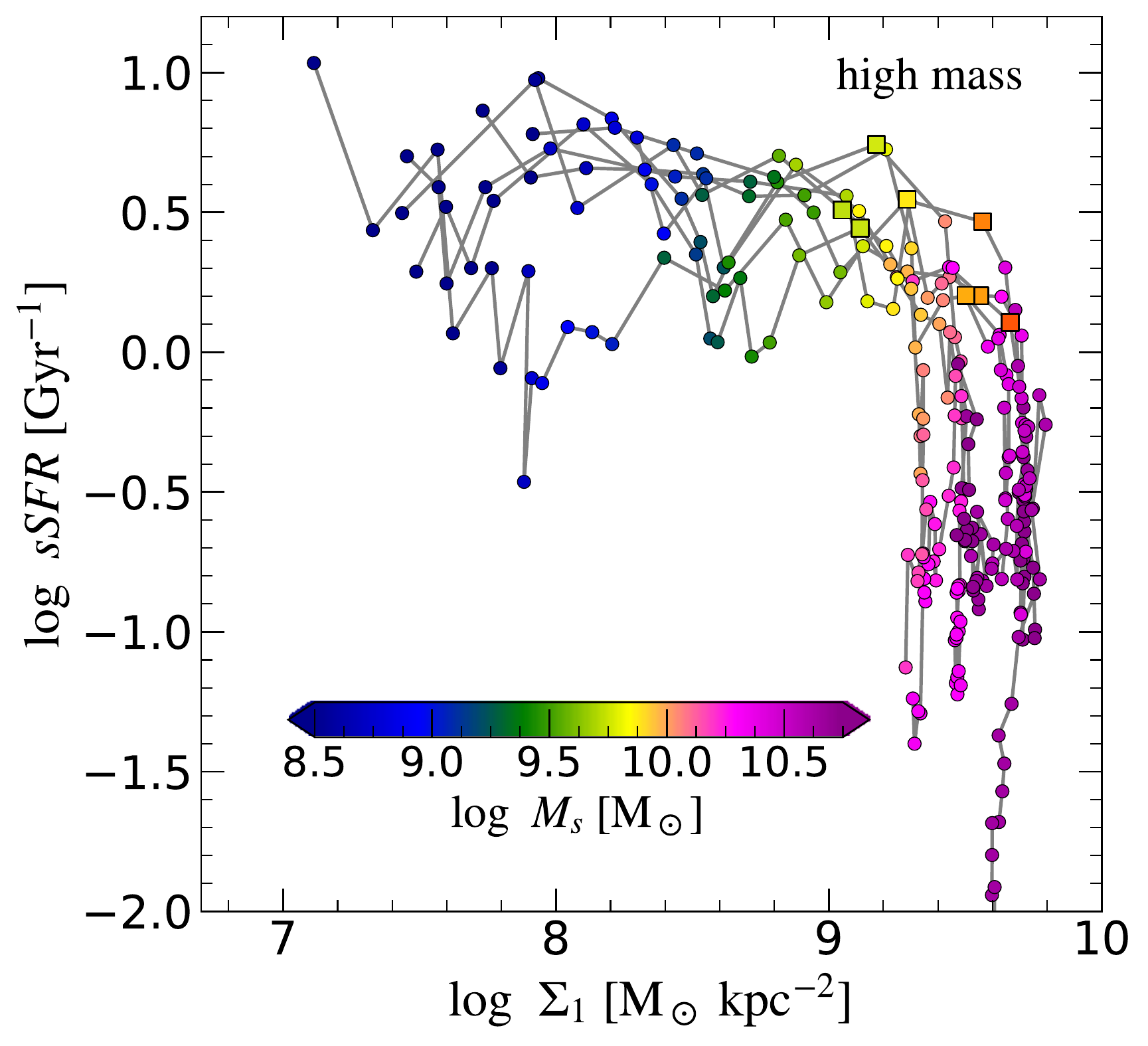}
\caption{
Wet compaction into a ``blue nugget" followed by quenching, as
seen in cosmological simulations.
{\bf Left:}
A schematic diagram of a wet-compaction event (corresponding pictures in
\fig{mosaic_H2}),
presenting the time evolution of masses (left axis in $\log (M/\msun)$)
inside the inner $1\kpc$ \citep[following][]{zolotov15}.
The wet compaction is driven by the sharp increase in gas mass (blue), 
by an order of magnitude
during a period of about $0.3\, t_{\rm Hubble}$. It rises to a peak as a blue
nugget (BN), and quickly declines by central-gas depletion into 
star formation and outflows while there is no replenishment.
The SFR (green, right axis in $\log (\msun \yr^{-1})$) reflects the gas mass,
showing post-compaction quenching in the inner $1\kpc$.
The stellar mass in the center (red) is rising following the boosted SFR
during the compaction process, flattening off after the blue nugget phase.
The central $1\kpc$ is dominated by dark matter (black) prior to the compaction
and by baryons (mostly stars, red) after the compaction.
The kinematics is dominated by dispersion pre compaction and by
rotation post compaction. 
The major blue-nugget peak typically occurs when the galaxy is near the golden
mass, $\Ms \sim 10^{10}\msun$, between the earlier supernova-dominated
phase and the post-compaction phase of a hot CGM.
The black-hole mass growth (times $10^4$, dashed black) 
is suppressed by supernova feedback below the golden mass, pre-compaction,
and it becomes rapid during and after the compaction in
the hot-CGM phase above the golden mass. The rapid black-hole
growth is triggered by the compaction process \citep[see also][]{dekel19_gold}.
{\bf Right:}
The universal evolution track for eight simulated galaxies (VELA simulations), 
relating sSFR and $\Sigma_1$, the inner stellar surface density inside $1\kpc$,
serving as a measure of compactness
(Lapiner, Dekel et al., in preparation).
The compactness is increasing while the sSFR is roughly constant (horizontal
tracks)
during the compaction process, and the evolution turns over at the blue-nugget 
``knee" (a square symbol) to quenching while 
$\Sigma_1$ remains roughly constant (vertical tracks).
Observations show a very similar behavior \citep[][Fig.~7]{barro17},
where $\Sigma_1$ in the vertical tracks slightly increasing with 
redshift. The quenching here is not caused by AGN feedback, as it is not
incorporated in the VELA simulations.
}
\label{fig:compaction}
\end{figure*}

\begin{figure*} 
\centering
\includegraphics[width=0.9\textwidth]
{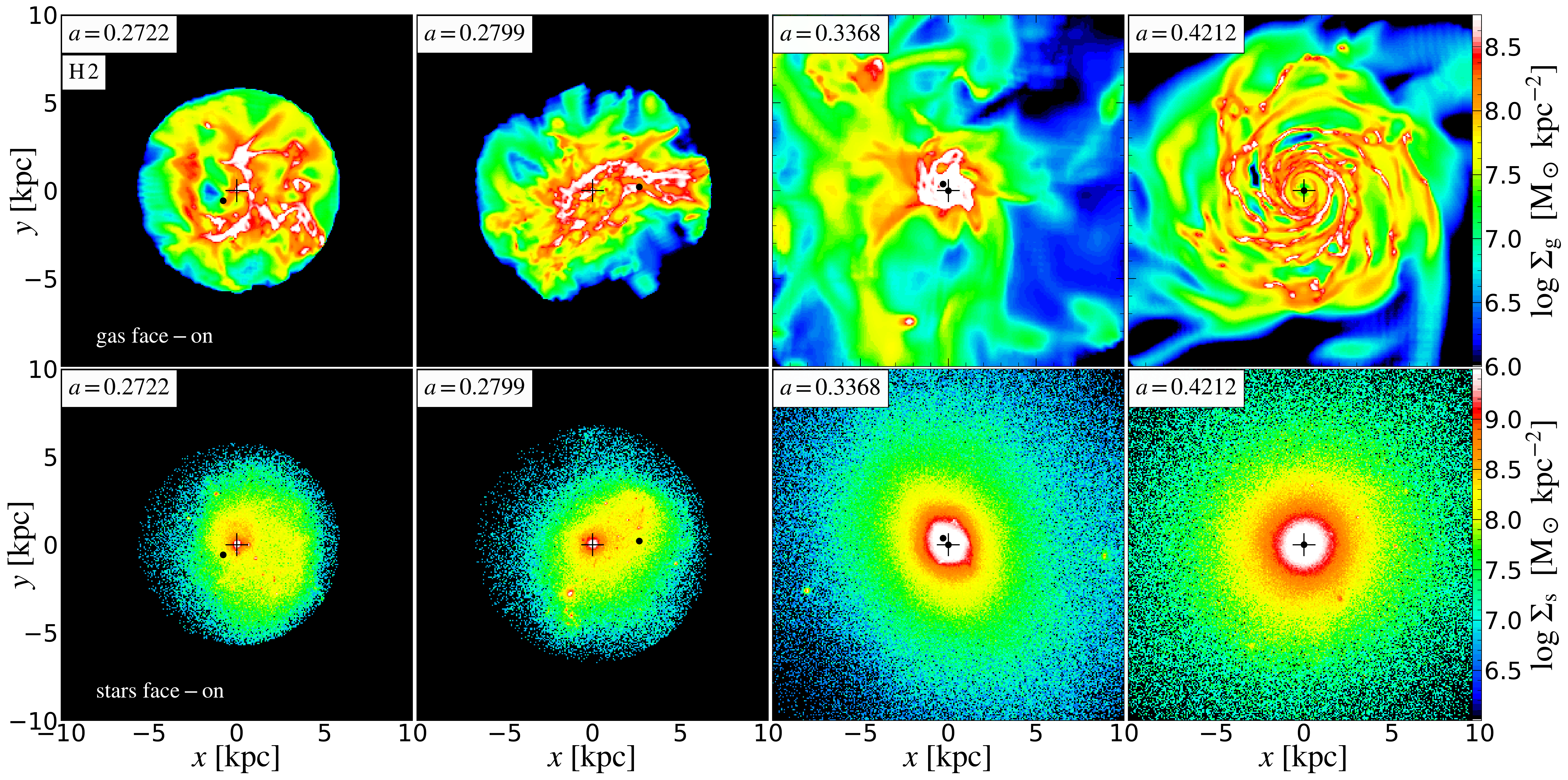}
\caption{
Wet compaction and black-hole sinkage to the center.
Shown are the projected densities of gas (top) and stars (bottom)
in different phases [the expansion factor $a=(1+z)^{-1}$ is marked]
during the evolution of one of the {NewHorizon} simulated galaxies (H2).
The simulations are to be presented in \se{NH} and in Appendix \se{app_NH}.
The projections are face on with respect to the gas angular momentum.
{\bf Top}, from left to right.
First and second: in the pre-compaction phase 
($\log \Ms \seq 9.8$). 
Third: at the blue-nugget phase ($\log \Ms \seq 10.9$). 
Forth: post-compaction VDI disc ($\log \Ms \seq 11.0$). 
{\bf Bottom:} 
The stellar compact nugget forms during and soon after the
compaction and the resulting bulge remains compact thereafter.
The black-hole, marked by a filled circle,
is orbiting at $1\sdash 2\kpc$ about the center
in the pre-compaction phases.
It sinks to the center at the blue-nugget phase and
remains locked to the center thereafter.  
}
\label{fig:mosaic_H2}
\end{figure*}

We have learned from cosmological simulations that most galaxies undergo
a substantial
wet-compaction event, which typically occurs in a pronounced way when the
galaxy mass grows above a threshold at a golden value, 
$\Mv\!\sim\! 10^{12}\msun$ and $\Ms\!\sim\! 10^{10}\msun$,
This occurs especially at $z\!=\!1\!-\!5$ when the gas fraction tends to be
high.
The pioneering exploration of this sequence of events has been performed using
the {VELA} cosmological simulations
\citep{zolotov15,tacchella16_ms,tomassetti16,dekel19_gold,dekel20_ring}.
The wet-compaction process is a significant gaseous shrinkage into a compact 
star-forming system within the central $1\kpc$ -
a blue nugget.
The resultant gas consumption into new stars and the associated gas removal
by stellar and supernova feedback cause gas depletion from the center which
results in an inside-out quenching of star-formation rate (SFR)
\citep{tacchella16_prof}.

\smallskip
The left panel of \fig{compaction} illustrates the main 
characteristics of this succession of events
via the time evolution of gas mass, stellar mass and SFR within the inner 
kiloparsec, as seen in the simulations.
More directly comparable to observations, 
The right panel of
\fig{compaction} presents simulated evolution tracks of eight {VELA}
galaxies in the plane of two observable quantities, the specific SFR (sSFR) 
and $\Sigma_1$, the measure of compactness via 
the stellar surface density inside $1\kpc$.
The L-shape evolution track consists of a 
compaction at a roughly constant sSFR 
and a subsequent quenching at a constant $\Sigma_1$, 
with the ``shoulder" at the blue-nugget peak of compaction.
This universal L-shape evolution track has been confirmed observationally
\citep[e.g.][Fig.~7]{barro17}.
%
The choice of using a fixed $1\kpc$ as the characteristic radius of the nuggets 
for all galaxies is motivated by the empirical finding, both in the 
simulations and the observations, that it provides a robust universal behavior 
with tighter relations, e.g., in the evolution tracks of \fig{compaction}.
One obtains qualitatively similar results when using instead
the effective radius, that is typically on a comparable scale of $\sim\!1\kpc$
for galaxies in the range of masses and redshifts concerned, but with a
somewhat larger scatter.

\smallskip
\Fig{mosaic_H2} illustrates through images of gas and stellar surface
density the evolution through the compaction, blue-nugget and post-blue-nugget
phases in an example galaxy from the {NewHorizon} simulation.
Similar sequences of images are shown in \se{app_images} 
for two other {NewHorizon} galaxies, and for an example galaxy from the 
{VELA} simulations. 
They show a similar robust evolution pattern through the 
compaction event, despite the different numerical codes and the different
sub-grid physical recipes adopted in the two suits of simulations, including
the presence and absence of AGN feedback.

\smallskip 
From the observational perspective, 
the exploration of blue nuggets started with the discovery of abundant, 
compact, massive, passive galaxies
at $z\!\sim\! 2\!-\!3$, typically encompassing $\sim\! 10^{10}\msun$ of 
stars within $1\kpc$, termed ``red nuggets"
\citep{dokkum08,damjanov09,newman10,
dokkum10,damjanov11,whitaker12,bruce12,dokkum14,dokkum15}.
Their effective radii are about one percent of their halo virial radii,
smaller than expected had the gas started in the halo with a common spin 
parameter $\lambda \!\sim\! 0.035$ and conserved angular momentum (AM)
during its contraction \citep{fe80}.
This indicated dissipative contraction accompanied with AM loss, termed a wet
compaction \citep{db14}, implied the existence of gaseous blue nuggets 
as the direct progenitors of the red nuggets.
Indeed, the predicted blue nuggets have been observed soon thereafter, showing
properties in agreement with being the red-nugget progenitors
\citep{barro13,barro14_bn_rn,barro14_kin,williams14,barro15_kin,dokkum15,
williams15,barro16_alma,barro16_kin,barro17_alma,barro17}.
A deep-learning analysis, trained on mock dusty images of the blue
nuggets from the simulations, identified with high confidence
blue nuggets of similar properties in the multi-color CANDELS-HST imaging 
survey of galaxies at $z\!=\!1\!-\!3$ \citep{huertas18}.

\smallskip 
The loss of AM that leads to compaction is indicated in the simulations
to be triggered by gas-rich mergers ($\sim\! 40$ per cent),
by colliding counter-rotating streams, by recycling fountains
or by other processes,
and to be possibly associated with violent disc instability \citep{db14}.
These processes preferentially occur at high redshifts,
where the overall accretion is at a higher rate and richer in gas,
leading to deeper compactions into smaller radii with higher densities.

\smallskip 
The main compaction events induce major transitions in the galaxy
structural, compositional and kinematic properties
\citep[e.g.][]{zolotov15,tacchella16_prof,tacchella16_ms}.
The compaction generates inside-out quenching of star formation.
This is accompanied by a transition from an irregular and diffuse 
configuration to a compact body, to be surrounded by an
extended gas-rich ring \citep{dekel20_ring}
and possibly embedded in a stellar envelope.
Kinematically, the system evolves from being pressure supported 
to rotation supported.
As a result of the compaction, the central region turns from being dominated
by dark-matter to being dominated by baryons.
This induces a global-shape transition of the stellar system 
from prolate to oblate 
\citep{ceverino15_shape,tomassetti16},
in agreement with observations \citep{vanderwel14_shape,zhang19}.

\smallskip 
For our current purpose, it is important to realize that the blue nuggets tend
to form near a characteristic mass.
Minor compaction events are seen in the simulations to occur at all masses
during the history of a star-forming galaxy (SFG).
Repeated episodes of minor compactions followed by quenching
attempts can explain the confinement of SFGs to a narrow Main Sequence
\citep{tacchella16_ms}.
On the other hand, the major compaction events, the ones that involve 
an order-of-magnitude increase in central density, cause a transition from
dark-matter  to baryon dominance and trigger deep quenching,
occur near the golden mass of $\Mv\!\sim\!10^{11.5-12}\msun$,
as can be seen in 
\citet[][Fig.~8]{tomassetti16} and \citet[][Fig.~21]{zolotov15}.
A deep-learning study of {VELA} simulated galaxies against
observed CANDELS galaxies \citep{huertas18}, indeed 
confirmed a favored stellar mass for the observed blue nuggets
near $\Ms \!\sim\! 10^{9.5-10}\msun$.
The presence of a preferred mass for blue nuggets makes the compaction-induced 
transitions of galaxy properties appear as pronounced systematic variations 
with mass near the golden mass.

\smallskip 
Our understanding is that
the major compaction events tend to be confined to near the
golden mass primarily because supernova feedback, which suppresses the
compactions at lower masses, becomes ineffective near and above the golden mass 
\citep{dekel19_gold}.
Supernova feedback at the central regions, which is boosted following the 
starburst soon after the early phases of the compaction process,
removes central gas and suppresses further compaction in
galaxies below the critical mass for efficient supernova feedback.
Supporting evidence for the effect of supernova feedback on the depth of the
compaction events is provided by simulations with stronger feedback, which show
compaction events that are less dramatic (e.g., the {NIHAO} and
{VELA6} simulations, in preparation).
Near and above the golden mass, where the potential well is deep enough,
with an escape velocity above the supernova-driven wind velocity (\se{sn}),
and when it becomes even deeper due to the compaction itself, 
the compaction is not significantly affected by supernova feedback 
and the contraction can proceed to higher central densities. 

\smallskip 
The characteristic mass for compaction events, being in the ball park of the
golden mass of galaxy formation and the indicated mass threshold for 
rapid black-hole growth, may indicate that the compaction
events have a major role in the transition from slow to rapid black-hole
growth, to be explored next.

\begin{figure*} 
\centering
\includegraphics[width=1.00\textwidth]{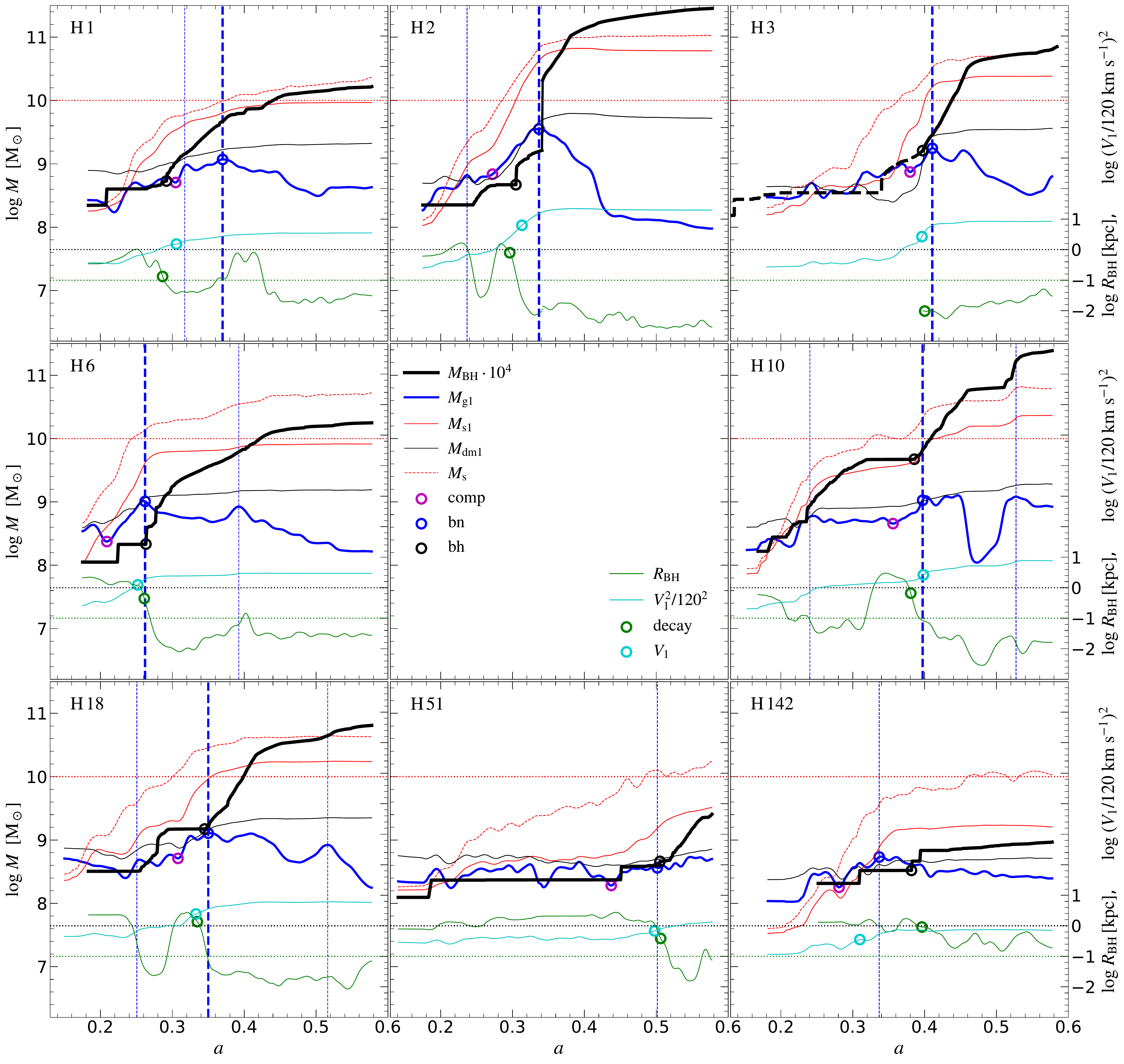}
\caption{
Compaction-driven black-hole growth. 
Shown are the evolution tracks in eight {NewHorizon} simulated galaxies
that reached $10^{10}\msun$ by $z \ssim 1$, as a function of cosmological
expansion factor $a=(1+z)^{-1}$ (where the time in the EdS regime, 
at $a \!<\! 0.5$, is $t\!\simeq\! 17.5\Gyr\, a^{3/2}$).
The main features to focus on are the 
evolution of black-hole mass ($\times 10^4$, black) and gas mass
within $1\kpc$ (blue).
The black-hole growth is suppressed by supernova feedback at early times,
when the galaxy is below the golden mass of $\Ms\!\sim\!10^{10}\msun$
(horizontal red line).
A major compaction event, marked by the rise of the gas mass to a peak at the
blue-nugget phase (vertical dashed blue), 
is identified in six of the eight galaxies, and seems to
trigger a rapid black-hole growth.
Minor compaction events are marked (vertical thin blue).
The blue nugget is also characterized by a rise to a plateau of the
stellar mass within $1\kpc$ (solid red) and the transition from central
dark-matter dominance (thin black) to baryon dominance, as in \fig{compaction}.
Also marked in some cases is a secondary compaction event (vertical 
thin dashed blue).
The total galaxy stellar mass (dashed red) at the blue-nugget phase 
is near the golden mass.
Also displayed is the position of the black-hole with respect to the galaxy 
center (green, $\log R$, right axis), showing that 
pre compaction the black-hole is wandering about the galaxy center at 
$\ssim 1\kpc$ (horizontal black line, right axis),
while the compaction typically brings it to well inside the $100\pc$ vicinity 
of the center (horizontal green line, right axis).
This is a result of the deepening of the potential well (cyan, $\log V^2$,
right axis) and the increased drag.
Symbols mark five of the events that are associated with the compaction event 
and black-hole growth: 
the onset of compaction (``comp"), 
the peak of compaction at the blue-nugget phase (``bn"),
the onset of rapid black-hole growth (``bh"), 
the deepening of the potential well (``$V_1$"), 
and the sinking of the black-hole to the center (``decay").  
}
\label{fig:Mbh}
\end{figure*}

\begin{figure*} 
\centering
\includegraphics[width=1.00\textwidth]
{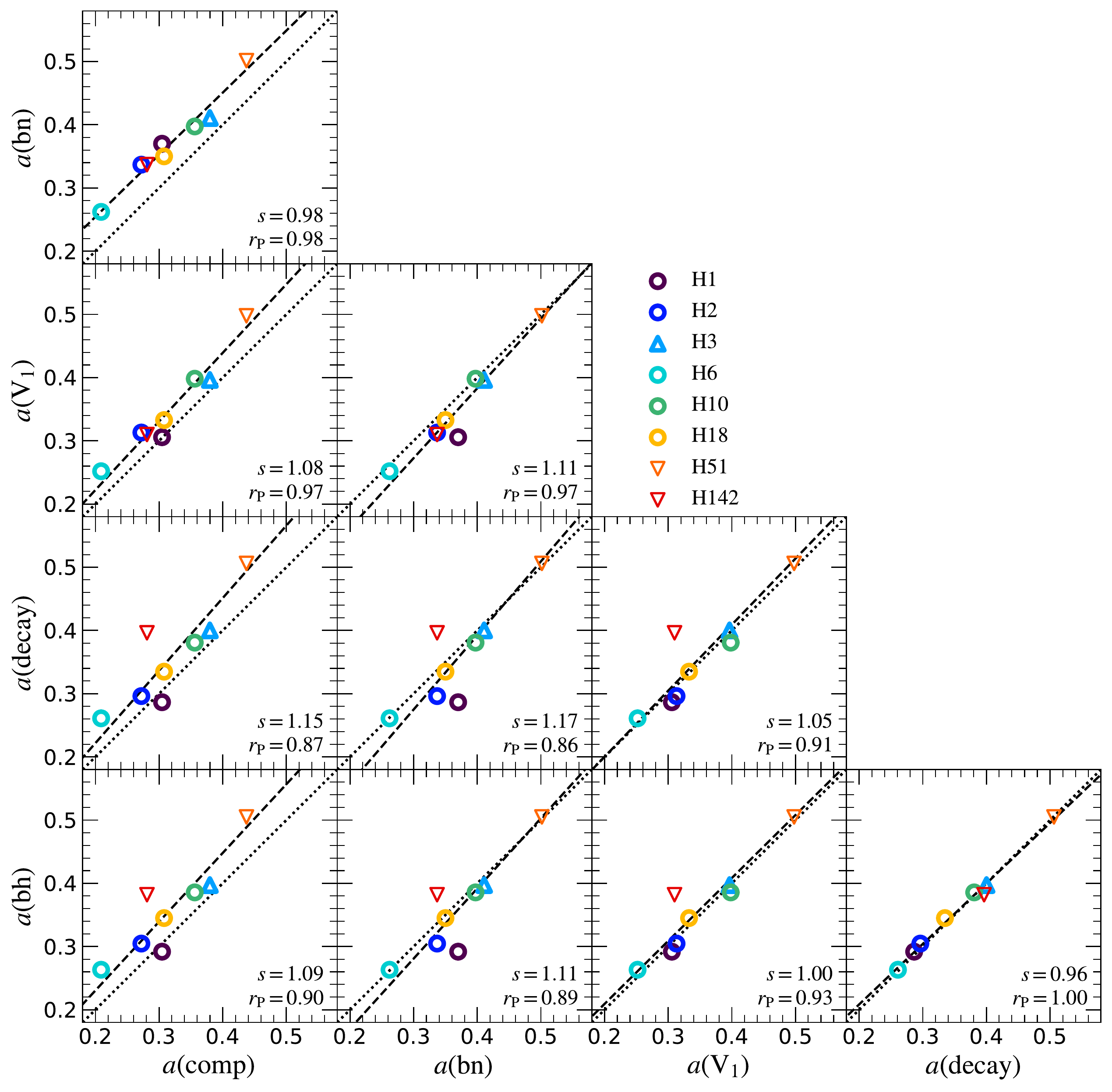}
\caption{
Correlations between the times of the five events associated with the 
compaction and the black-hole growth.
The symbols refer to the events in the eight {NewHorizon} galaxies, 
as marked in \fig{Mbh}.
The events are 
(a) the onset of rapid black-hole growth (bh),
(b) the onset of compaction (comp),
(c) the end of compaction at the blue-nugget (bn),
(d) the half-way up of the potential at $1\kpc$ ($V_1^2$),
and (e) the half-way down of the radius of the black-hole orbit (decay).
The dotted line is the identity relation and the dashed line is the 2D 
linear regression line.
The slope of the line and the Pearson correlation coefficient are marked by
``$s$" and $r_{\rm P}$ respectively. 
All the events are fairly tightly correlated, with a correlation coefficient 
$r_{\rm P}\!\sim\! 0.87\sdash 1.00$.
The small offsets in some of the correlations indicate that
the events associated with the BH growth tend to occur slightly after the
onset of compaction and slightly before the blue-nugget peak.
}
\label{fig:corr}
\end{figure*}

\section{Compaction and Black-Hole Growth}
\label{sec:bh}

We will demonstrate here, using simulations, that
the evolution of galaxies from the supernova zone below the golden mass,
through a compaction event near the golden mass, into the hot-CGM phase above
the golden mass, determines the black-hole growth rate and imprints the golden
mass of galaxies in it.
One expects that
below the golden mass, namely pre-compaction, the supernova-driven gas
heating and ejection from the center suppresses the black-hole growth.
Then,
the major compaction that tends to occur near the golden mass, where
supernova feedback is already inefficient, brings gas into the inner
sub-kiloparsec blue nugget. This may lock the black-hole into the galaxy 
center and induce efficient accretion onto the sub-parsec black-hole,
which can trigger a rapid black-hole growth and the activation of an AGN.
In more massive haloes, the deep potential well and the hot CGM are expected to
lock the central gas in, and allow continuing accretion onto the black-hole.

\smallskip 
The suppression of black-hole growth by supernova feedback below the golden
mass and the rapid black-hole growth above it have been seen in several
different cosmological simulations, using different codes and sub-grid models
and especially different implementations of supernova feedback and black-hole
growth.
This was first seen in a {RAMSES} simulation by \citet{dubois15}.
By comparing the same simulations with and without supernova feedback,
they clearly demonstrated that the suppression of black-hole growth below 
the golden mass is primarily due to supernova feedback.
Similar results were seen in {RAMSES} simulations by \citet{habouzit17},
in EAGLE simulations \citep{bower17}, in
FIRE simulations \citep{angles17} and in Illustris TNG \citep{habouzit18}.
\citep[also][]{prieto17, mcalpine18, trebitsch18, trebitsch20} %
In particular, \citet{bower17} discuss physical arguments for the suppression
of black-hole growth below the golden mass and its inefficiency above the 
golden mass in terms of the buoyancy of supernova bubbles into a colder CGM
versus a hotter CGM in the two zones respectively.

\subsection{Following Black Holes and Blue Nuggets in {NewHorizon}}
\label{sec:NH}

Our current analysis is based on the {NewHorizon} cosmological simulation
\citep{dubois20_nh}\footnote{https://new.horizon-simulation.org/science.html},
also used in \citet{park19_1,park19_2}.
This is a {RAMSES} AMR simulation \citep{teyssier02},
zooming in with a high maximum spatial resolution of $\sim\!40\pc$ 
($27\sdash 54\pc$) on a 
spherical volume of comoving radius $10\Mpc$, smoothly embedded in the 
larger box of the {Horizon-AGN} simulation
\citep{dubois14_nhagn,dubois16}.
By $z\seq 0.7$, this volume, encompassing a loose group of galaxies,
has eight galaxies that have grown above the golden mass of 
$\Ms \ssim 10^{10}\msun$ by that time, which we use in our current analysis.
The assumed $\Lambda$CDM cosmological parameters resemble the WMAP-7 results
\citep{komatsu11},
namely,
flat curvature, $\Omega_{\rm m}\seq 0.272$, $\Omega_{\Lambda}\seq 0.728$,
$\Omega_{\rm b}\seq 0.045$, $\sigma_8\seq 0.81$, $h_0\seq 0.704$, 
and $n_{\rm s}\seq0.967$.

\smallskip 
The sub-grid physics incorporates both supernova feedback and black holes with
AGN feedback, as described in \se{app_NH} based on \citet{dubois20_nh}
and the {NewHorizon} website. We bring here a  
a brief summary of relevant features.
The mechanical supernova feedback models successively the phases of energy 
and momentum conservation \citep{kimm14,kimm15}, 
and considers the effect of clustered supernovae
\citep{kim17,gentry17,gentry19}, 
as well as the momentum due to pre-heating near OB stars \citep{geen15}.
Massive black holes are implemented as sink particles with a seed mass of
$10^4\rm\, M_\odot$ and zero spin, 
which form in cells where both gas and stellar density exceed the 
density threshold for star formation, and away from neighboring black holes.
The black-hole is growing by the Bondi-Hoyle gas accretion rate
\citep{bondi44}, modulated by a spin-dependent radiative efficiency that
is computed on-the-fly, and limited by the Eddington limit.
The required averaged gas properties in the vicinity of the black-hole are
measured within a radius of $\sim\! 150\pc$ \citep{dubois12}, in which 
the maximum resolution of $\sim\! 40\pc$ is kept at all times.
AGN feedback is modeled in two modes with different efficiencies in the regimes
of high and low accretion rate,
determined by the ratio of accretion-to-Eddington rate.
In the \emph{"radio mode"}, at low accretion rates,  
the AGN feedback is in the form of bipolar jets aligned with the black-hole
spin. 
In the \emph{"quasar mode"}, at high accretion rates,
the AGN is assumed to deposit thermal energy within a sphere
of radius comparable to one cell, with an efficiency that is
coupled to the spin-dependent radiative efficiency
\citep{dubois12}.
The black-hole spin is followed on-the-fly due 
to gas accretion and black-hole mergers,
which occur when the pair separation is $<\!150\pc$
and the relative velocity is lower than the escape velocity of
the binary.

\subsection{Supernova suppression and compaction-driven growth}
\label{sec:compaction_bh}

\Fig{Mbh} shows the evolution of black-hole mass (black curve, 
multiplied by $10^4$)
as a function of expansion factor $a\seq(1+z)^{-1}$ in the eight 
{NewHorizon} 
simulated galaxies that have grown to above $\Ms \seq 10^{10}\msun$.
Shown in comparison are the masses of gas, stars and dark matter  
within the inner $1\kpc$, as well as the total stellar mass.
The gas inside $1\kpc$ (blue) provides the clearest indication for a 
wet-compaction event, where the gas density rises from the onset of compaction
to a blue-nugget peak, followed by gas depletion into star formation 
and partly into weak supernova-driven outflows that may still persist near the 
golden mass.
This is similar to the generic evolution pattern derived from the {VELA}
simulations shown in the cartoon, \fig{compaction}, and for several simulated
galaxies in \citet[][Fig.~2]{zolotov15}.
The onset of compaction and the blue-nugget peak, as crudely identified by eye,
are marked. The compaction typically lasts for  
one third of the cosmological time during the event
(as in the EdS cosmological regime, valid at $z\!>\!1$, 
$\Delta t/t \ssim (3/2) \Delta a/a$, and $\Delta a/a \ssim 0.2$). 
During the compaction, the central gas density increases by a factor 
$3\sdash 10$.
The history of SFR within the inner $1\kpc$ (not shown) follows closely
the history of the central gas density in accord with the Kennicutt-Schmidt 
relation \citep{kennicutt98,zolotov15,dubois20_nh}. 
During the major compaction event, 
the associated stellar mass within $1\kpc$ (red) rises from the onset of
compaction until the blue-nugget peak and it flattens off to a constant 
value soon thereafter, a typical increase by a factor of ten.
This shoulder in the central stellar-mass curve distinguishes the major 
compaction to the blue-nugget SFR peak from other possible peaks that mark
additional minor compaction events \citep{tacchella16_ms}. 
The total stellar mass (dashed red) at the blue-nugget peak
is typically $\sim 10^{10}\msun$, reflecting the golden stellar mass scale,
as seen in the {VELA} simulations and discussed in \se{compaction}.

\smallskip 
In all cases, we see a suppressed black-hole growth in the pre-compaction,
low-mass, supernova regime, followed by a rapid growth post-compaction, into 
the high-mass, hot-CGM regime, consistent with earlier simulations. 
The impression is that the rapid black-hole growth 
starts soon after the onset of compaction, and it continues throughout the 
compaction process and later on into the hot-CGM regime.
Galaxies H2, H6 and H18 provide the clearest cases, where both the compaction 
and the onset of black-hole growth are pretty well defined.
In H1 
there is an earlier minor compaction (near $a\seq 0.24$),
which is not easily
separated from the major compaction, making the identification of the
onset of the major compaction uncertain. 
In H3, the black-hole mass prior to $a\seq 0.4$ is marked dashed because the
black-hole used to reside in another massive galaxy that merged with H3 near 
that time (where compaction occurred and the black-hole started to grow 
near $a\seq 0.34$). 
Missing the full information about the other galaxy (for a
technical reason), the association of galaxy properties and black-hole growth 
in H3 prior to $a\seq 0.4$ is uncertain.
In H10, we identify minor compactions both before and after the major
compaction, near $a\seq 0.24$ and $0.53$.
In the lower-mass galaxy H51, in which the onset of rapid
black-hole growth occurs later, shows only a marginal gas compaction event 
near that time, which is distinguished mostly by the growth of central
stellar density that has not reached the plateau by $z\seq0.7$.
Finally, in H142, the compaction is relatively well defined, though it is
rather weak.

\smallskip 
The correlation between the onset of black-hole growth, at $a({\rm bh})$,
and the compaction event, via its onset and blue-nugget peak at 
$a({\rm comp})$ and $a({\rm bn})$,
respectively, is shown in the two left panels of the bottom row in \fig{corr}.
The onset of rapid black-hole growth is defined as the upturn to a continuous
long-term steep growth of black-hole mass.
If a black-hole merger occurs just prior to this upturn, the onset of
black-hole growth is identified with the black-hole merger event.
We see that
the onset of black-hole growth tends to occur slightly after the onset of 
compaction and slightly before the blue-nugget phase. 
A symmetric linear regression yields a fairly tight correlation, 
with a Pearson correlation coefficient of $r_{\rm p} \ssim 0.9$, 
and a slope of $s \ssim\! 1.1$, not far from unity. This is a fairly convincing
correlation, given the large uncertainties in the visual identifications of the
events.

\smallskip 
The impression from the evolution of central gas mass in \fig{Mbh} is that
the major compaction events in {NewHorizon} are 
qualitatively similar to the ones in {VELA}, despite 
the presence of AGN feedback only in the former.                               
This indicates that AGN feedback does not have a major role in driving the
compaction process.
Furthermore, one can see in \fig{Mbh} and \fig{corr} that
the onset of rapid black-hole growth and its sinkage to the center (see below)
tend to occur slightly after the onset of compaction, consistent with a causal
relation where the compaction triggers the rapid black-hole growth, 
which activates strong AGN feedback.

\subsection{Black-hole wandering and lockdown}

\subsubsection{In the simulations}

A clue for how the compaction actually triggers the black-hole growth is 
provided by tracing the position of the black-hole with respect to the galaxy 
center, shown as the green curve in \fig{Mbh}.
One can see that during the pre-compaction phase the
black-hole tends to orbit at one to a few kiloparsecs off the center
(which is by itself not always accurately defined in this irregular phase), 
where the orbital
velocity and the supernova-driven dilute-gas environment naturally
suppress accretion onto the black-hole. 
This off-center wandering is also seen in the images shown in \fig{mosaic_H2} 
and in \se{app_images}.
Similar wandering of the black holes off the centers has been
seen in simulations of low-mass galaxies \citep{bellovary19,pfister19}. %
As seen in \fig{Mbh} and in \fig{mosaic_H2},
during and soon after the compaction, the black-hole sinks to inside the
$\sim 10-100\pc$ vicinity of the galaxy center, corresponding to a couple
of cell sizes, and stays there.
This can be interpreted as being due to the compaction-driven deepening of 
the potential well, shown in \fig{Mbh} as the cyan curve marking $V_1^2$, 
with $V_1$ the circular velocity at $1\kpc$,
as well as the drag against the compaction-driven dense baryons and dark
matter.
Being locked to the gas-rich galaxy center, with only little gas removal by 
supernova feedback above the golden mass, 
the black-hole is now subject to efficient accretion.

\smallskip 
\Fig{corr} also shows the correlations between the events discussed
in \se{compaction_bh} (namely, the onset of black-hole growth, the onset 
of compaction, and the blue-nugget phase) 
and the events associated with the sinkage of the black-hole  
to the center, namely, the time $a(V_1)$ when $V_1$ is half way up from its
pre-compaction value to its post-compaction value, and the time of decay 
$a({\rm decay})$ when the radius of the black-hole orbit is half way down its
pre-compaction value.
These events are also marked on the evolution tracks of each galaxy in
\fig{Mbh}.
We note that the value of $V_1$ at the transition is in the ball park of
$V_{\rm SN} \ssim 120 \kms$, marking the critical potential-well depth for 
effective supernova feedback.
We see good correlations with negligible offsets of  
$a({\rm bh})$ with $a({\rm decay})$ and $a(V_1)$.

\subsubsection{Analytic estimates}

Why is the black-hole wandering off center when the galaxy is below the 
golden mass?
We recall that since $V_1 \slt V_{\rm SN}$ in this regime, supernova feedback
can remove the gas from the inner $1\kpc$, which suppresses the 
black-hole growth. 
The shallow potential well, in a low-mass halo and with dark matter but only 
little baryons at the center, allows the black-hole to wander.
The black-hole may be pushed away from the center partly because in this mass 
range the mergers are more frequent than an orbital time, 
$t_{\rm merg} \slt t_{\rm orb}$ \citep{dekel20_flip}. 
The mergers may increase the orbital angular momentum of the black-hole,
and can generate clumps off which the black-hole
can scatter \citep[see][]{pfister19}. %
Furthermore, the mergers can trigger bursts of star formation which, through
supernovae, could drive turbulence in the remaining central gas and  
push it to larger radii.
As we show next, the dynamical friction (DF) that is exerted on the black-hole
by the dark matter and gas, which in principle could decrease the orbital AM of 
the black-hole and pull it to the center, is weak at this stage.

\smallskip 
For a black-hole in a circular orbit at radius $r$ with velocity $V^2(r) = 
GM(r)/r$, 
the timescale for losing all its AM by DF deceleration $f_{\rm df}$ can be
estimated by the Chandrasekhar approximation \citep{chandrasekhar43}, 
\be
t_{\rm df} \simeq \frac{V}{f_{\rm df}} \simeq
\frac{1}{4\,\pi\, \ln \Lambda \, G^2} \frac{V^3}{\rho\, M_{\rm bh}} \, ,
\label{eq:tdf}
\ee
where $\rho$ is the background density of particles with velocities smaller
than $V$ and $\Lambda$ is the Coulomb parameter that 
under certain assumptions %
could be crudely 
approximated by $\Lambda \ssim \Mv/M_{\rm bh}$ such that $\ln\Lambda \ssim 10$.
With the orbital time $t_{\rm orb} \seq 2\,\pi\, r/V$ (which is $\sim\! 30\Myr$
for $10^{10}\msun$ inside $r\seq 1\kpc$),
and with $M(r) \!\simeq\! (4\pi/3)\, \rho\, r^3$,
we obtain 
\be
t_{\rm df} \sim 50\, \left(\frac{M(r)}{M_{\rm bh}}\right)_4 \,t_{\rm orb}\, ,
\label{eq:tdf2}
\ee
where the mass ratio is in units of $10^4$.
Pre-compaction, when the black-hole growth is suppressed, one can see in 
\fig{Mbh}
that at $r \ssim 1\kpc$ the mass ratio is significantly larger than $10^4$, 
implying that the DF is ineffective for a few hundred orbital times, 
which is in the ball park of a Hubble time.

\begin{figure*} 
\centering
\includegraphics[width=1.00\textwidth]{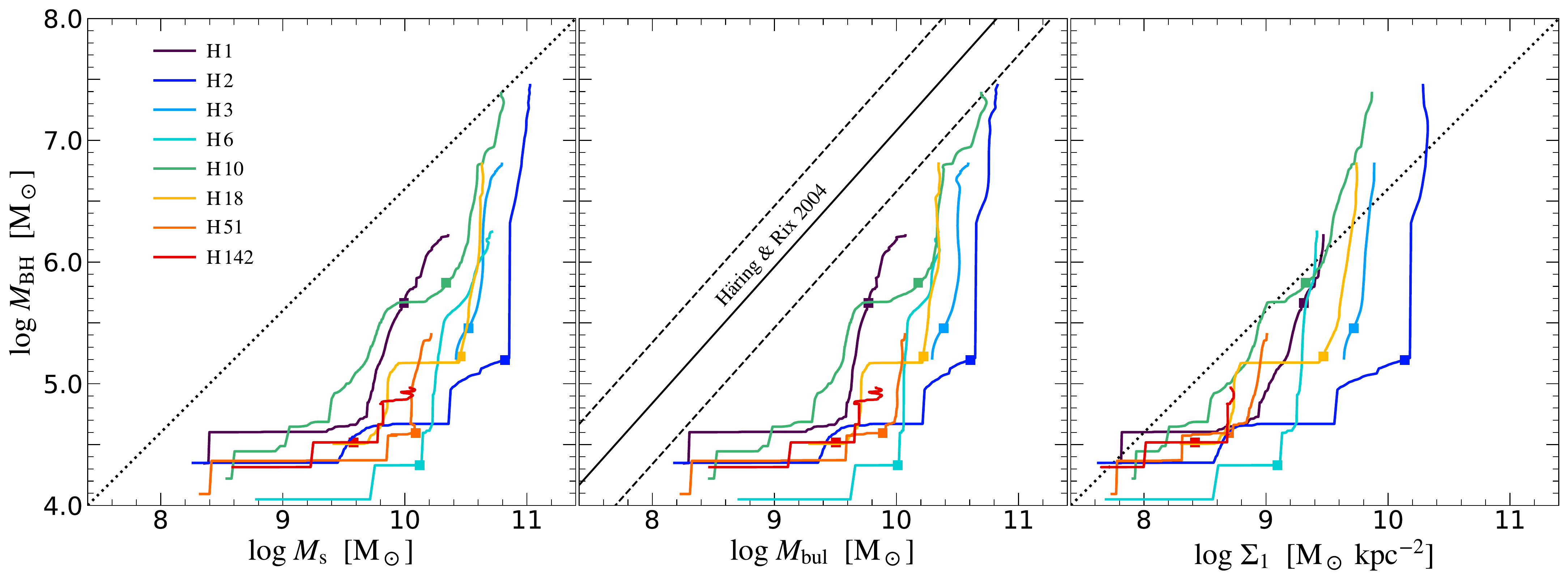}
\caption{
Compaction-driven black-hole growth in the eight {NewHorizon} galaxies.
The evolution tracks of black-hole mass versus stellar mass $\Ms$
show (supernova-driven) suppression of black-hole growth below
$\Ms\!\sim\!10^{10}\msun$, where $\Sigma_1\!<\!9 \msun \kpc^{-2}$,
turning into a rapid growth near the critical mass,
likely driven by compaction events (squares).
Black holes of $\sim\!10^{5}\msun$ in galaxies of $\Ms\!\sim\!10^{9.5}\msun$
are predicted to lie below the standard linear relation.
}
\label{fig:Mbh-Ms}
\end{figure*}

\begin{figure} 
\centering
\includegraphics[width=0.40\textwidth]{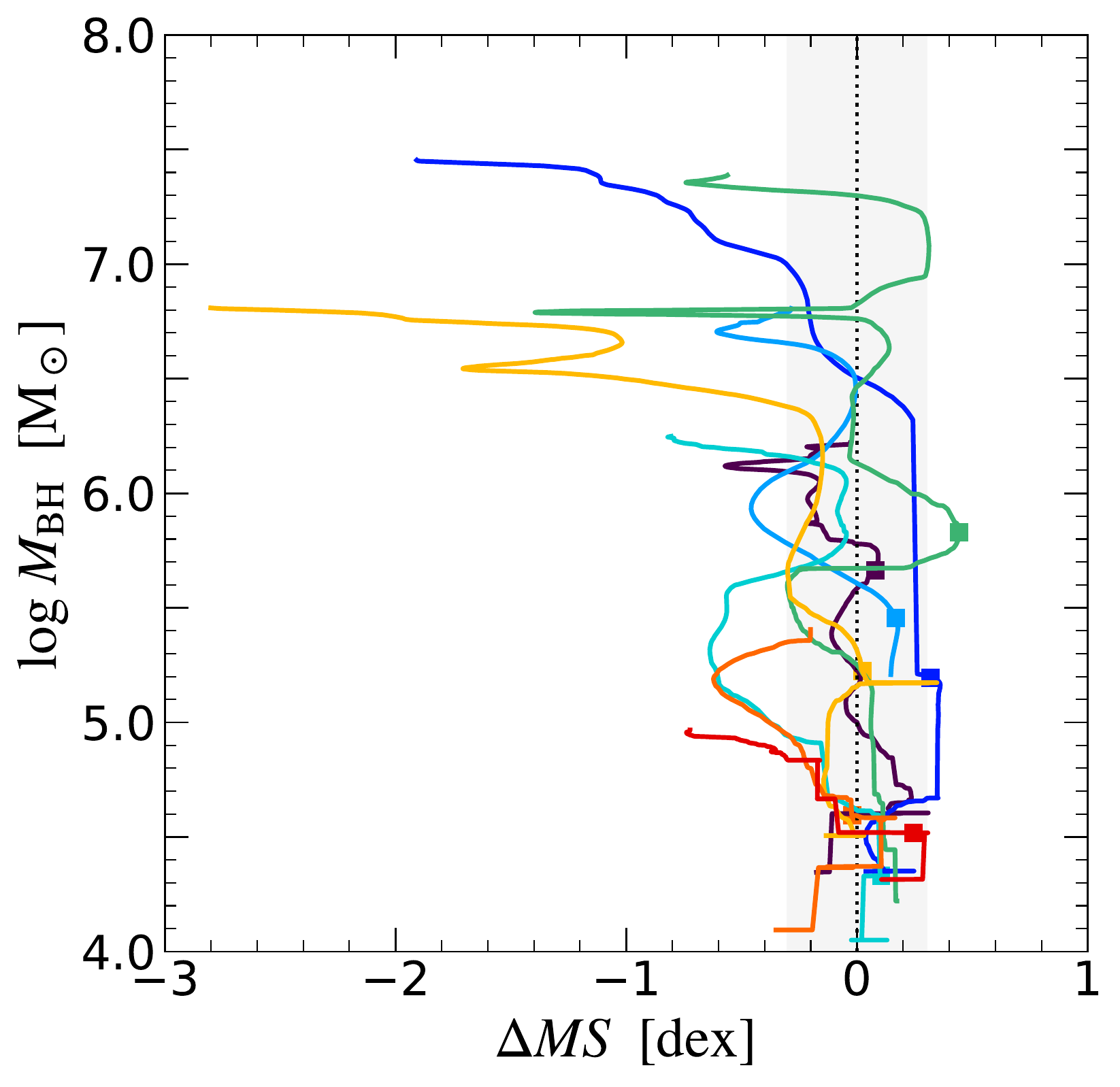}
\caption{
Black-hole mass versus deviation of sSFR from the ridge of the
star-forming Main Sequence (MS) at the given redshift
in the eight {NewHorizon} simulated galaxies.
The $\pm 0.3$dex MS is marked by the shaded area.
The blue nuggets are marked by squares.
Much of the black-hole growth occurs on the MS and some in the Green Valley
at negative $\Delta$MS values.
}
\label{fig:Mbh-dms}
\end{figure}

\begin{figure} 
\centering
\includegraphics[width=0.47\textwidth]{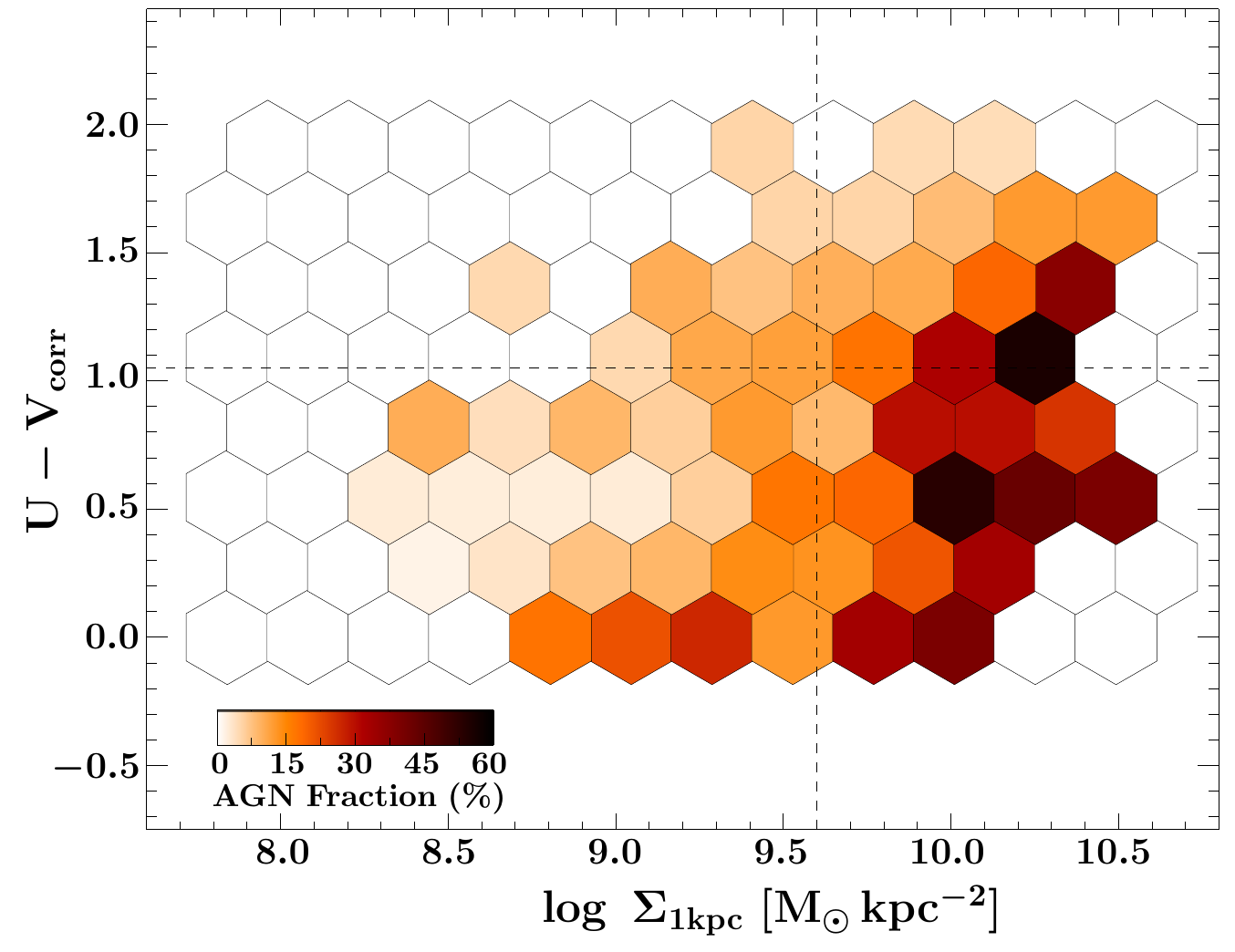}
\caption{
Fraction of CANDELS galaxies that host X-ray detected AGN (color scale)
in the plane of SFR (growing downwards) and compactness,
as represented by U-V color and stellar surface
density inside $1\kpc$ $\Sigma_{\rm 1\kpc}$ respectively
\citep[following][]{kocevski17}.
The AGN fraction is high in the blue-nugget quadrant of compact star-forming
galaxies (bottom-right), consistent with compaction-driven BH growth.
}
\label{fig:kocevski}
\end{figure}

\smallskip 
What is the effect of the compaction-driven deepening of the potential well?
As can be seen in \fig{Mbh},
the compaction near the golden mass makes the mass within $1\kpc$ grow by a
factor $f \ssim 10$, causing a contraction of the black-hole orbit.
This occurs over a period $\Delta a \ssim 0.05$, which corresponds 
at $a\seq 0.25$ to $\Delta t \!\simeq\! 26 \Gyr\, a^{1/2} \Delta a$
namely $\Delta t \ssim 650\Myr$.
Thus, the potential at $1\kpc$ changes on a timescale much longer than the 
orbital time of a few Myr, making the black-hole orbit contract 
adiabatically by a factor 
$f \ssim 10$ \citep{flores93}\footnote{A contraction by a factor of two is 
obtained in the other limit when the compaction is instantaneous 
\citep{freundlich19_udg}}.

\smallskip 
Once above the golden mass, the time between mergers becomes longer than the
orbital time, so the pushing out of the black-hole
becomes weaker and it can continue
to sink to the center.
With $V_1 \sgt \Vsn$ in this regime, there is no significant gas removal 
by supernova feedback, and the slowly moving black-hole is free to accrete from 
the compaction-driven enhanced gas density and grow in mass.

\smallskip 
The black-hole growth and its sinkage to the center post-compaction 
is a runaway process, because, according to \equ{tdf}, $t_{\rm df}$ is 
rapidly decreasing as $M_{\rm bh}$ is growing and as the black-hole moves 
inwards where $V(r)$ is smaller and $\rho(r)$ is larger.
With $M(r)/M_{\rm bh} \ssim 10^3$ in \equ{tdf2}, we have 
$t_{\rm df} \ssim 150\Myr$, which corresponds to a shrinkage of the black-hole
to the
center over $\Delta a \sim 0.01$, consistent with the steep shrinkage seen 
in \fig{Mbh}. 
The specific Bondi-Hoyle
accretion onto the black-hole is similarly growing in 
proportion to the black-hole mass and the background density, 
$\dot{M}_{\rm bh}/M \ssim 4\pi G^2 \rho M_{\rm bh} / v^3$, 
where $v$ is the root-mean-square of the velocity dispersion and the 
sound speed in the accreting region about the black-hole.

\smallskip 
We conclude that the compaction locks the black-hole to the galaxy center 
due to the combination of adiabatic contraction by the deepening of the
potential well and the dynamical friction that is boosted by the increase in
density and black-hole mass. 
Indeed, we see in \fig{corr} that the onset of black-hole growth
is correlated with the sinkage of the black-hole to the center, 
which is correlated with the deepening of the central potential well.

\smallskip 
We do not know to what extent the black-hole growth would have been suppressed
in the supernova regime had the black-hole been artificially locked to the
galaxy center. We note that the raw curves for gas mass within $1\kpc$, 
before they were smoothed as in \fig{Mbh}, show significant fluctuations 
between output timesteps (\fig{Mbh_raw}). 
This could be interpreted as episodic removal of gas by supernovae, 
which may indicate that the black-hole growth would have been suppressed even 
if locked to the center, possibly implying that the wandering of the black 
hole off center is only one reason for the suppression of growth. 
 
\begin{figure*} 
\centering
\includegraphics[width=0.46\textwidth]{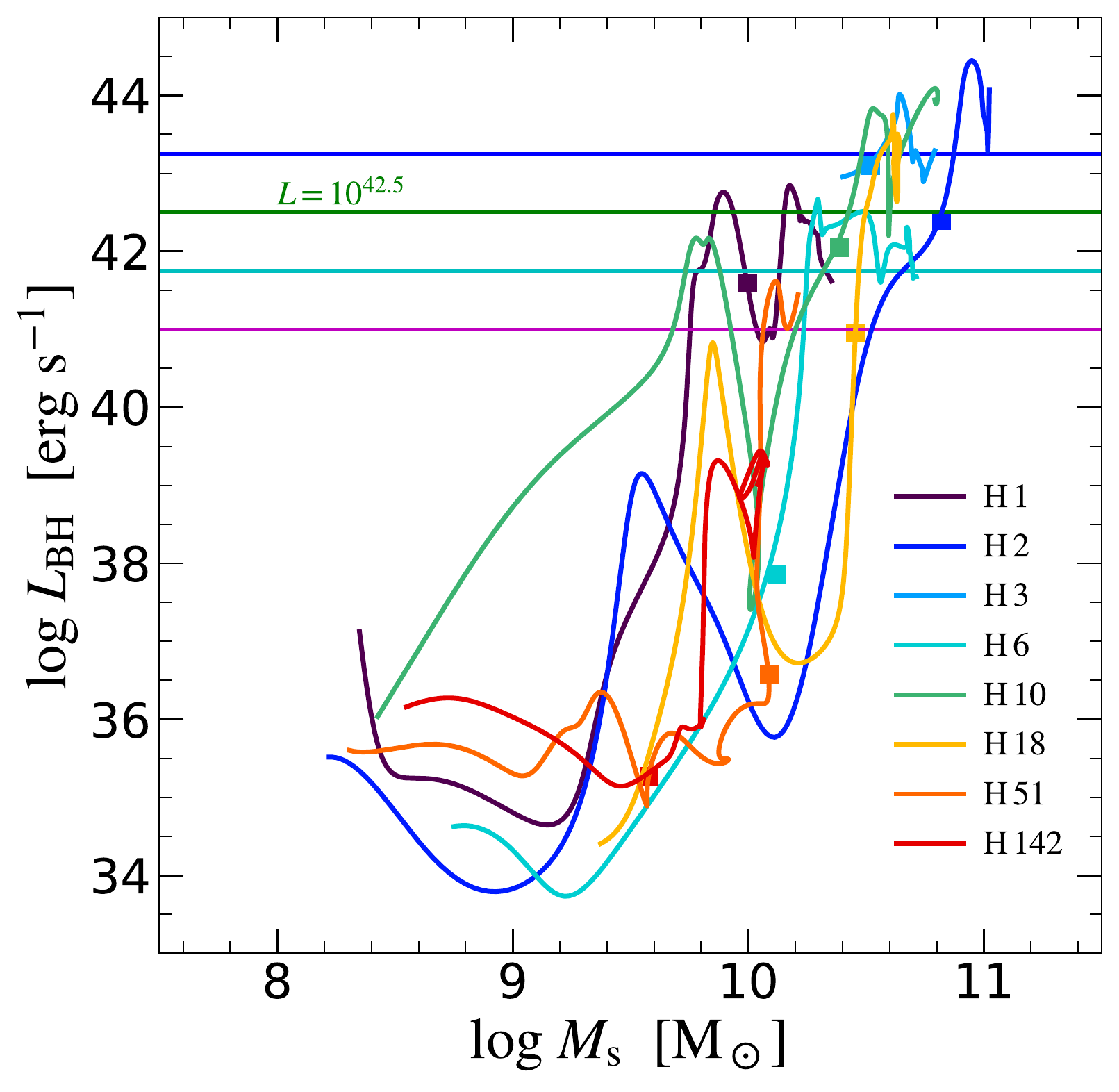}
\includegraphics[width=0.48\textwidth]{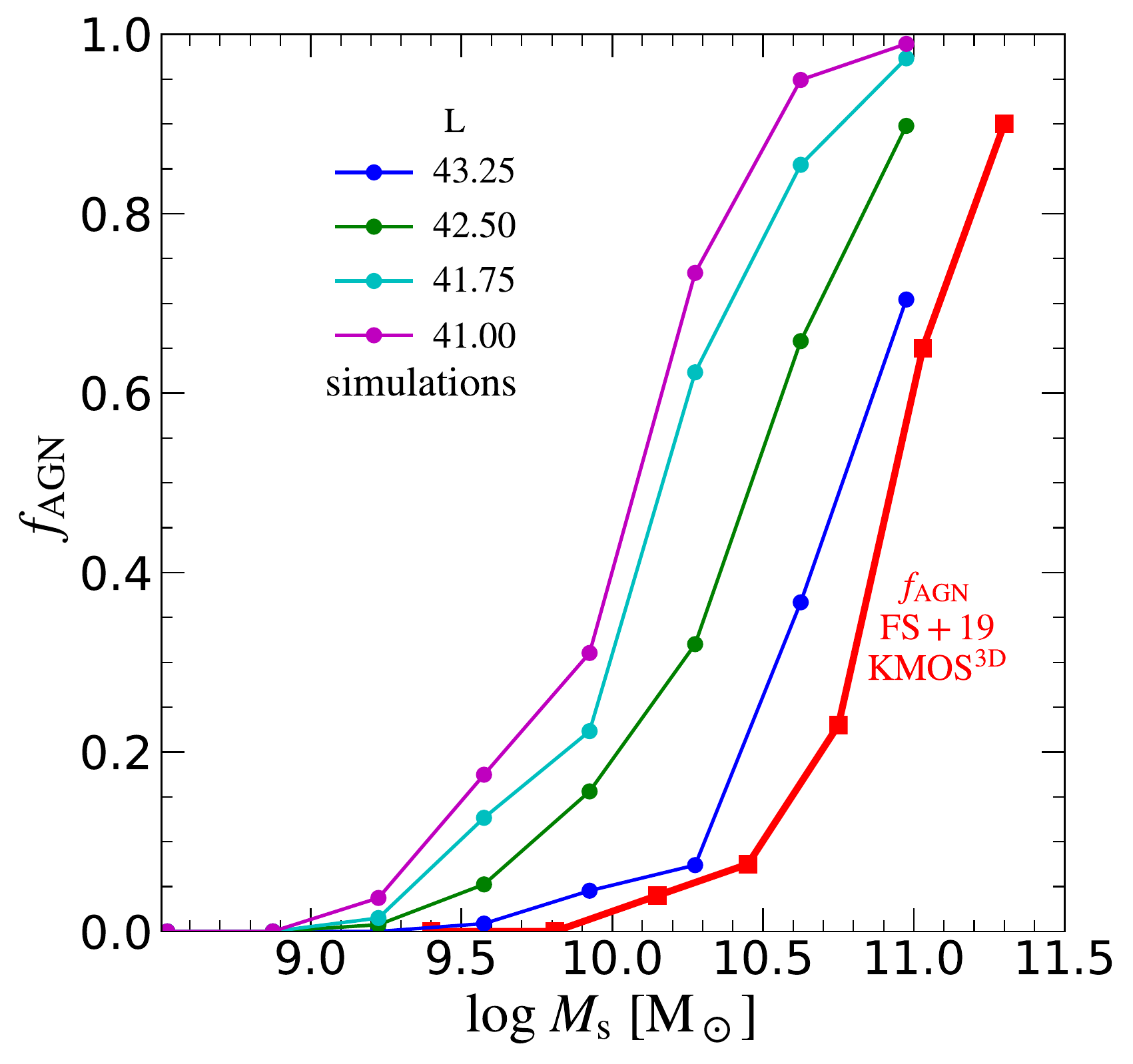}
\caption{
{\bf Left:}
AGN luminosity versus stellar mass in {NewHorizon}
simulated galaxies.
The luminosity is derived from the accretion rate onto the black-hole,
excluding BH mergers, using the spin-dependent radiative efficiency, \equ{L}.
We see a general L-shape track with a turn-up near the golden mass,
analogous to \fig{Mbh-Ms}.  The blue-nuggets are marked by squares.
{\bf Right:}
When AGN are selected above a luminosity threshold (lines in the left panel),
the fraction of galaxies that host AGN is plotted versus $\Ms$.
The observed AGN fraction adapted from \citet{forster19}
is marked by the red curve, where the threshold
luminosity is about $10^{42.5}\ergs$.
The L-shape curve is reproduced by the simulations, with a slight offset toward
lower masses or higher AGN fractions.
}
\label{fig:Mdotbh}
\end{figure*}

\subsection{Scaling relations of black holes and galaxies}

\Fig{Mbh-Ms}
puts together the eight simulated galaxies, showing the evolution
tracks of black-hole mass versus total stellar mass $\Ms$,
bulge mass, and the stellar surface density within the inner
$1\kpc$, the latter serving as our most effective measure of compactness.
A characteristic L-shape evolution is seen, displaying a suppression of 
black-hole growth below the golden mass of $\Ms \ssim 10^{10}\msun$
and a rapid black-hole growth once above the golden mass.
The blue-nugget phase for each galaxy, marked by a square symbol,
typically coincides with the turn-up of the curve, consistent with a causal
connection between the two.
A similar turn-up is found to occur at a characteristic
threshold of central surface density at $\Sigma_1\!\sim\!10^9 \msun \kpc^{-2}$,
consistent with the black-hole growth being driven by the compaction event.
Similar tracks are seen when $\Mbh$ is plotted against the bulge mass,
predicting that black holes of $\sim\! 10^5\msun$ should tend to lie below
the standard linear relation between black-hole mass and bulge mass that is
observed to be valid for black holes more massive than $10^6\msun$  
\citep[e.g.][]{magorrian98,haring04,kormendy13,ho14}.

\smallskip  
\Fig{Mbh-dms} 
shows for the same eight galaxies the black-hole mass versus the deviation 
$\Delta$MS of the log SFR from the ridge of the Main Sequence of 
star-forming galaxies, as it varies with time.
We use for the ridge sSFR$_{\rm MS}\seq 0.04 (1+z)^{5/2}$ \citep{dekel13},
which is indeed a good fit to the {NewHorizon} SFGs.
This figure may allow an exploration of the relationship between the black-hole 
growth and the evolution of the galaxies along the MS and the eventual
beginning of the quenching process through the Green Valley at negative
$\Delta$MS values.
During the pre-compaction and the compaction phase we find confined 
oscillations about the Main-Sequence ridge \citep{tacchella16_ms},
with the blue-nugget peak typically above the ridge.
Rapid black-hole growth occurs while the galaxy is still on the Main sequence.
Three galaxies in the sample (H2, H10, H18) eventually drop well below
the Main Sequence, when the black-hole mass is above $\sim\! 10^{6.5}\msun$
(with H10 experiencing a rejuvenation episode by a late compaction event).
This may allow testing of the hypothesis put forward by \citet{chen20} 
that much of the black-hole growth occurs in the Green Valley.
Some growth occurs at negative values of $\Delta$MS, toward the upper Green
Valley, but the onset of rapid black-hole growth seems to precede the entry 
to the Green Valley.

\subsection{AGN fraction}

\Fig{kocevski}, following \citet{kocevski17},
shows the fraction of CANDELS-survey galaxies that host X-ray detected AGN in
the plane of SFR, represented by U-V color, versus compactness,
as measured by the
stellar surface brightness within $1\kpc$, $\Sigma_{1}$ [compared to
$\Sigma_{\rm e}$ in \citet{kocevski17}].
This plane is similar to the plane shown in \fig{compaction} where the L-shape
evolution tracks are detected in the simulations, 
except that here the SFR is growing downwards.
One can indeed see a high AGN fraction in the quadrant of high SFR and high
compactness representing the blue nugget phase of evolution, in accordance with
what we see in the simulations.
Similar evidence comes from obscured AGN that tend to reside in compact 
star-forming galaxies \citep{chang17}.

\smallskip
We next attempt to reproduce the fraction of galaxies that host AGN as a 
function of their stellar mass, in order to compare to observations such
as \citet[][Figs. 6,13]{forster19}.
The left panel of \fig{Mdotbh} shows the AGN luminosity as a function of
stellar mass for the eight massive NewHorizon galaxies shown in the previous
figures.
The luminosity is derived from the gas accretion rate onto the black-hole,
averaged over fine timesteps (median $\Delta t$ of $\sim\! 0.3\Myr$),
excluding black-hole growth by mergers.
The growth rate is derived in the simulations from the Bondi-Hoyle accretion rate
\citep{bondi44}
by $\dotMbh\!=\!(1\!-\!\epsr)\,\dot{M}_{\rm Bondi}$,
where $\epsr$ is the radiative efficiency of the black-hole based on its spin
as calculated on-the-fly in the simulations.
The luminosity is then given by
\be
L = \frac{\epsr}{1-\epsr}\, \frac{d\Mbh}{dt}\, c^2 \, .
\label{eq:L}
\ee
The radiative efficiency tends to grow with mass, 
where on average $\epsr/(1-\epsr) \!\simeq\! 0.25$.
Using this average as a constant value in \equ{L} does not make a qualitative 
difference to the results.
For the sake of clarity, the luminosity curves in the left panel of 
\fig{Mdotbh} were smoothed over $\sim\!120\Myr$.

\smallskip
The resultant AGN fraction is shown in the right panel of \fig{Mdotbh},
for each of four different luminosity thresholds in the selection. 
These are compared to the red curve which represents
the fraction observed by \citet{forster19}, who report to have
used an effective luminosity threshold of $10^{42.5}\ergs$.
The predicted curves all resemble in their L shape the observed curve, with the
turn-up mass or the AGN fraction shifted as a function of the chosen 
luminosity threshold.
When assuming in the simulations the same threshold as estimated for 
the observed 
sample (green), the turn-up mass is apparently underestimated by about half
a decade,
while a better fit is provided by the simulations with a brighter
threshold of $10^{43.25}\ergs$ (blue).
This offset may reflect an uncertainty in the effective luminosity threshold 
applied in the observations or in the assumed correspondence of luminosity to 
accretion-rate in the simulations.
It may also be a result of a real underestimate of the golden mass in the 
simulations or an overestimate of the simulated accretion rate. 
In regard with possibly underestimating the golden mass, 
one should note that while the existence of
a golden mass is robust, rooted in the physical phenomena discussed above,
the exact value of the golden mass in the simulations may be subject to the 
way the sub-grid physics of supernova feedback and black-hole growth are 
implemented.

\smallskip
\citet{forster19} also provide (their figure 9) 
an indication for an anti-correlation between
the AGN fraction and the stellar effective radius for galaxies of
$\Ms \ssim 10^{10.00-10.75}\msun$, consistent with the predicted
compaction-driven AGN activity at and above the golden mass.
Their figure 3 indicates a correlation of AGN fraction with $\Delta$MS, the
deviation from the ridge of the Main Sequence (MS) of star-forming galaxies at
a given stellar mass, consistent with the proposed onset of AGN activity at the
blue-nugget phase, where the galaxy is above the MS ridge
\citep{tacchella16_ms}, similar to what is seen observationally 
in \fig{kocevski}.

\section{Quenching Trigger \& Maintenance}
\label{sec:quenching}

The gained understanding of how black-hole growth is triggered by a compaction 
event
has implications on the SFR quenching of massive galaxies and in particular
the possible role of AGN feedback in triggering the quenching and in
maintaining it for long term.

\smallskip
In order to understand the evolution within the central $1\kpc$ of a galaxy,
and in particular the quenching process at and above the golden mass,
it is helpful to appeal to the key parameter $\tinf/\tdep$,
the ratio of timescales for gas inflow into this region and
gas depletion from it, by consumption to star formation and by 
outflows \citep{tacchella16_ms}.
Wet compaction is possible if $\tinf<\tdep$, such that significant dissipative
compaction can occur before the gas turns to stars or be ejected by
feedback \citep{db14}.
This is also the condition for halting an ongoing depletion-quenching event and
rejuvenating a new compaction event by newly accreted gas.
On the other hand, the condition for quenching is $\tdep<\tinf$, ensuring that
significant depletion occurs before the gas may be replenished by accretion.

\smallskip
In the cold-flow regime below the critical mass for virial shock heating,
the specific inflow rate into the galaxy roughly follows that of the
cosmological total specific accretion rate into the halo.
In the Einstein-deSitter cosmological regime (roughly valid at $z\!>\!1$), 
this is derived analytically and confirmed by
simulations \citep{dekel13} to be
\be
\tinf \sim 25 \Gyr\, (1+z)^{-5/2}\, M_{12}^{0.14} \, ,
\ee
where $M_{12}$ is the halo mass in $10^{12}\msun$.
It implies that below the golden mass $\tinf$ is a strong function of
redshift and a weak function of mass.
The inflow time becomes much longer once the halo is above the golden 
mass, where the heated CGM suppresses the cold gas supply,
and especially at $z\! <\! 2$, when the penetration of cold streams through the
hot CGM is suppressed \citep[\fig{scale_z}][]{db06,cattaneo06}.

\begin{figure} 
\centering
\includegraphics[width=0.477\textwidth]{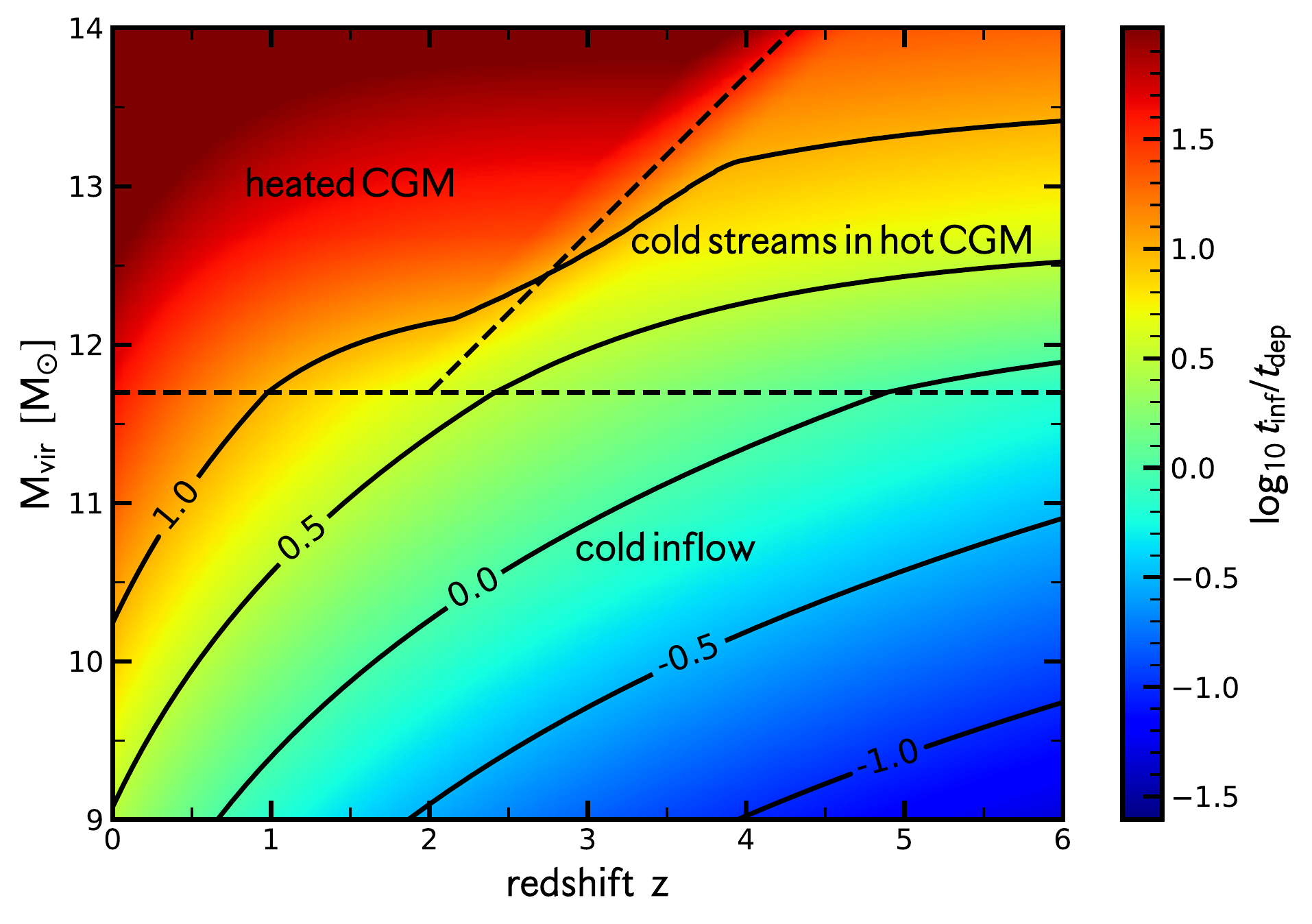}
\caption{
The ratio of timescales for gas inflow-replenishment and for depletion from the
inner $1\kpc$, in the $\Mv-z$ plane,
distinguishing between necessary conditions for wet compactions
and star formation when $\tinf/\tdep <1$ and deep quenching when
$\tinf/\tdep \!>\! 1$ \citep[as in][]{tacchella16_ms}.
Resembling the characteristic masses of \fig{scale_z},
the border line is near the golden mass at $z\! \leq\! 2$,
rising toward higher redshifts to allow for cold streams,
consistent with Fig.~13 of \citet{behroozi19} derived from observations.
}
\label{fig:tacchella}
\end{figure}

\smallskip
The average depletion time near the ridge of the Main Sequence of SFGs,
as estimated from the {VELA} cosmological simulations 
in which AGN feedback is not incorporated \citep{tacchella16_ms}, is 
\be
\tdep \sim 1 \Gyr\, (1+z)^{-0.5} \, M_{12}^{-0.2}\, ,
\ee
namely a rather weak dependence on both mass and redshift.
This turns out to be a good approximation to the observational estimates
at $z \seq 1\sdash 3$ \citep{tacconi18}.
We thus have for masses below the golden mass
\be
\frac{\tinf}{\tdep} \sim 25\, (1+z)^{-2}\, M_{12}^{0.34} \, .
\label{eq:tinf_to_tdep}
\ee
For example, for galaxies of $M_{12} \ssim 0.3$, we have $\tinf \ssim \tdep$
at $z \ssim 3$.
At higher redshifts, it is more likely to have $\tinf/\tdep \!<\!1$,
namely a necessary condition for wet compaction once there is a trigger,
followed by a burst of star formation.
At lower redshifts, where $\tinf/\tdep \!>\!1$ is more common, 
there is an efficient post-compaction central gas depletion with no efficient 
replenishment by fresh cold gas, allowing deeper long-term quenching, 
especially when the CGM is hot above the golden mass.

\smallskip
\Fig{tacchella}
shows the expected quenching efficiency, the ratio $\tinf/\tdep$, as a
function of halo mass and redshift, based on \equ{tinf_to_tdep} 
\citep[see also][]{tacchella16_ms}.
There is an encouraging qualitative resemblance between this and
Figure 13 of \citet{behroozi19}, which shows in the same plane the 
distributions of SFR and of the fraction of star-forming galaxies 
as estimated from observations via abundance matching with $\Lambda$CDM haloes.
We learn that our qualitative argument, which does not yet include the possible
effect of AGN feedback, does recover the quenching efficiency deduced from 
observations.
This is alongside with the success of the L-shape evolution track seen in
\fig{compaction}, also based on simulations with no AGN, in recovering the 
observed behavior for the onset of quenching \citep{barro17}.

\smallskip
Our picture of quenching is that,
overall, the quenching mechanism is primarily a function of mass.
In haloes of masses below the golden mass there is quenching by stellar and
supernova feedback. Especially at high redshifts, cold gas supply may cause a
new compaction event and a new burst of star formation, with boosted supernova
feedback that triggers a new quenching attempt. The associated oscillations of
$\tinf/\tdep$ about unity cause oscillations about the Main Sequence of SFGs,
which can explain the confinement of SFGs to a narrow Main Sequence
\citep{tacchella16_ms}.
Near the golden mass, a major compaction event triggers a central burst of
star formation, which leads to the onset of deeper quenching by gas consumption
and supernova feedback that is not followed by efficient replenishment, as 
$\tinf/\tdep$ becomes larger than unity.
The triggered quenching is maintained in haloes above the critical mass,
at least partly and for a while, because
the shock-heated CGM suppresses gas supply to the central galaxy. 
This is especially
efficient at low redshifts ($z<2$), where cold streams do not bring gas in
above the critical mass \citep{db06,bdn07}.
On the other hand, 
the hot gas in the inner halo can cool and ``rain" into the galaxy,
thus maintaining a non-negligible level of star formation. This is why the
quenching in hydro-cosmological simulations with no AGN feedback is 
incomplete in many cases,
as can be seen in \fig{compaction} \citep{zolotov15,tacchella16_ms}. 

\smallskip
However,
we have learned in the current paper that the major compaction event 
near and above the golden mass, in the hot-CGM regime,
generates a rapid growth of the central black-hole. 
Based on \fig{Mbh} and \fig{corr}, the onset of rapid black-hole growth indeed
tends to occur slightly after the onset of compaction.
The black-hole growth activates an energetic AGN that can help keeping the
CGM hot and thus maintain the quenching of star formation for a long time
\citep[preliminary analysis e.g. in][]
{croton06,bower06,cattaneo07,cattaneo09,dubois10,dubois16}. %
This can be tested by correlating the CGM temperature with the AGN activity,
which is beyond the scope of this paper.
This is consistent with the quenching by AGN after the end of the rapid
black-hole growth shown in Fig.~6 of \citet{dubois15}.
According to this scenario, AGN feedback is not the trigger of quenching.
It can rather serve as the major source for long-term maintenance of quenching 
in sufficiently massive hot haloes above the golden mass,
after the AGN has been triggered by a major wet-compaction event,
the same event that triggered the onset of quenching.

\section{Conclusions}
\label{sec:conc}

There is a rather robust observed threshold for luminous AGN at a 
characteristic mass
corresponding to dark-matter haloes of $\Mv \sim 10^{12}\msun$.
Not being aware of any obvious hint in black-hole physics for a 
characteristic mass of this sort,
we investigated how it could be imprinted on the central black-hole by the 
processes associated with galaxy evolution.
Indeed, observations reveal a golden mass scale for efficient star formation in
galaxies within dark-matter haloes at a comparable mass,
which also translates to the golden time for star formation at
$z \ssim 2$, when the mass of typical forming haloes 
is comparable to the golden mass.

\smallskip 
Two physical processes naturally confine the golden mass. On the low-mass side,
energy considerations imply that supernova feedback is effective in
suppressing star formation in haloes below the golden mass \citep{ds86}.
On the high-mass side,
an analysis of virial shock stability reveals that
the halo CGM is heated to the virial temperature in haloes above the golden
mass \citep{db06}.
The resemblance of the critical masses associated with these two different
processes (\fig{scale_z}) confines efficient galaxy formation to a
peak about the golden mass.
%
One may wonder whether this similarity between the two critical masses is a
coincidence or a built in match.
On one hand, the analyses in the two cases are related in the sense that the
preferred scale arises from comparing a radiative cooling time to the relevant
dynamical time. On the other hand, these timescales refer to different
environments on different scales. The conditions for cooling (e.g., gas
density and metallicity) and the relevant dynamical times 
(in star-forming clouds versus inflow into the halo) are very different in 
the two cases.

\smallskip 
We pointed out and demonstrated here that the connection of the golden mass 
to the black-hole growth becomes more concrete once we consider the 
wet-compaction events that
typically occur in the history of galaxies near the golden mass,
as shown by simulations \citep[here and in][]{zolotov15,tomassetti16} 
and as seen in observations
\citep[e.g.][]{barro13,barro17,huertas18}.
These events induce major transitions in galaxy properties,
where in particular they trigger the quenching of star formation by central
gas depletion into star formation and outflows.
The compaction processes are due to drastic angular-momentum losses,
about half caused by mergers and the rest by counter-rotating streams,
recycling fountains and other mechanisms.
The major deep compaction events to blue nuggets, those that trigger a decisive
long-term quenching process and a transition from central dark-matter to baryon
dominance,
tend to occur near and above the golden mass, at all redshifts.
We argue that this preferred mass scale for major compactions is due to the
same two physical processes of supernova feedback and hot CGM, which suppress
compaction attempts at lower masses and partly at significantly higher masses
\citep{tacchella16_ms}.

\smallskip 
We demonstrated using the {NewHorizon} simulation \citep{dubois20_nh}
that the compaction 
events naturally connect the characteristic mass for rapid black-hole growth
and AGN activation to the golden mass of galaxy formation.
The two zones, of supernova feedback and hot CGM, provide the necessary
conditions for black-hole suppression and rapid growth, respectively
\citep{ds86,dubois15,bower17}.
Black-hole growth is suppressed by supernova-driven gas removal
in star-forming galaxies of mass below the golden mass, while they
can grow above the golden mass as the gas is confined to the center
by the halo potential well and the hot CGM.
In between the supernova and CGM zones, we have demonstrated that
the onset of rapid black-hole growth is driven by the major compaction event
that tends to occur near the golden mass due to the same two physical 
processes.

\smallskip 
We found that the onset of black-hole growth in this simulation is 
associated with a sinkage of the black-hole from its orbit at $\sim\! 1\kpc$ 
to the center, due to the deepening of the potential
well and the increase of drag as a result of the compaction event.

\smallskip 
The quenching of star formation at and above the golden mass is thus proposed 
to be a multi-stage process.
It is triggered by a major compaction event, via central gas depletion due to
star formation and outflows.
The quenching is then maintained, at least partly and for a while, 
by the hot CGM that suppresses the cold gas
supply to galaxies in haloes above the golden mass, especially at $z\!<\!2$ 
when cold streams have hard times penetrating the hot CGM.
Finally, AGN feedback, activated by the compaction-driven black-hole growth,
can help keeping the CGM hot and thus maintaining long-term quenching.
We note that in this picture, contrary to commonly assumed scenarios,
AGN feedback is not responsible for the onset of quenching. Instead, these
two phenomena result from the same wet compaction event once the galaxy is 
near the golden mass defined by supernova feedback and the hot CGM. 
AGN feedback kicks in later on as a possibly important source for
quenching maintenance.

\smallskip 
It may be interesting to complement this picture by considering the 
general interplay between the galaxy and its environment
on large scales and its central region during the different phases.
In the star-forming supernova zone below the golden mass, cold flows that
originate
from outside the halo induce star formation that is regulated by supernova
feedback, which suppresses the central black-hole growth.
Near the golden mass, major compaction events that are induced externally by
mergers and counter-rotating streams form compact star-forming blue
nuggets, which trigger central quenching inside out. The compaction (outside-in)
causes a rapid central black-hole growth.
Above the golden mass, the hot CGM maintains the quenching by suppressing
the cold gas supply from the halo to the center.
The halo deep potential well and hot CGM lock the supernova ejecta and the
black-hole to the galaxy center and allow a continuous black-hole growth.
The resultant central AGN then helps the quenching on larger scales by 
keeping the CGM hot for a long time.

\section*{Acknowledgments}
We thank Dale Kocevski and Sandro Tacchella for assistance with figures 8 and 10.
We acknowledge stimulating discussions with Sandy Faber, Natascha 
Forster-Schreiber, Jonathan Freundlich, Katarina Kraljic, Reinhard Genzel 
and Joel Primack.
This research has been partly supported by 
BSF 2014-273, NSF AST-1405962,
GIF I-1341-303.7/2016, DIP STE1869/2-1 GE625/17-1
and ISF 861/20.
We acknowledge the {NewHorizon} simulation builders team for allowing the use 
of their non-public data, including Hoseung Choi, Julien Devriendt, 
Sugata Kaviraj, Katarina Kraljic, Taysun Kimm, 
Min-Jung Park, Sebastien Peirani, Christophe 
Pichon, Marta Volonteri and Sukyoung Yi.
The NewHorizon work was granted access to the HPC resources of CINES under  
allocations c2016047637, A0020407637 and A0070402192 by GENCI, 
KSC-2017-G2-0003 by KISTI, and as a ``Grand Challenge" project granted by 
GENCI on the AMD Rome extension of the Joliot Curie supercomputer at TGCC. 
This work has made use of the Horizon cluster on which the simulation was 
post-processed, hosted by the Institut d'Astrophysique de Paris.
We warmly thank S. Rouberol for running it smoothly.
\section*{Data availability}
The data underlying this article were provided by the {NewHorizon} collaboration. 
Data will be shared on request to the corresponding author with permission of the 
{NewHorizon} collaboration.

\bibliographystyle{mnras}
\bibliography{bh_comp_SL.bib} 

\appendix

\section{The {NewHorizon} Simulation}
\label{sec:app_NH}

\subsection{General}

The {NewHorizon} simulation \citep{dubois20_nh} is a hydro-gravitational
zoom-in cosmological simulation using an Eulerian adaptive mesh refinement 
{RAMSES} code \citep{teyssier02}. 
The gas is evolved using a second-order 
Godunov scheme to solve the Euler equations and an HLLC Rieman solver using 
the MinMod total variation diminishing scheme for the interpolation of cell 
quantities at the interface between cells.

\smallskip
The simulation is based on the standard $\Lambda$CDM cosmology 
with the cosmological parameters consistent with 
the {WMAP-7} data \citep{komatsu11}, namely flat curvature,
total matter density $\Omega_{\rm m}=0.272$, dark energy density 
$\Omega_{\Lambda}=0.728$, baryon density $\Omega_{\rm b}=0.045$, 
amplitude of matter power spectrum $\sigma_8=0.809$, 
spectral index $n_{\rm s}=0.967$,
and Hubble constant $\rm H_{\rm 0}=70.4\,km\,s^{-1}\,Mpc^{-1}$.

\smallskip
The {NewHorizon} volume is a high-resolution sphere of radius 
$\rm 10\,h^{-1}\,Mpc$ embedded in the larger, lower resolution 
{Horizon-AGN} box of side $L_{\rm box} = 100\rm\,h^{-1}\,Mpc$ 
\citep{dubois14_nhagn}.
The initial conditions in {Horizon-AGN} had $1024^3$ dark-matter 
particles, while in {NewHorizon} the dark-matter particle mass 
is $4^3$ times lower,
$M_{\rm DM} \seq 1.2\times 10^6\msun$.
The refinement within the zoom-in region is done using a quasi-Lagrangian 
strategy: a cell is refined if the total mass within it surpass 8 times the 
initial mass.
An extra level of refinement is added when the expansion factor is doubled 
(i.e. $a=0.1, 0.2, 0.4, 0.8$), thus keeping the minimal cell size within 
the limits of $\Delta x=27-54\,\rm pc$.
In addition a super-Lagrangian refinement is added to the mesh if the cell 
size is below one Jeans length 
and 
where the gas number density exceed $5\rm\,H\,cm^{-3}$.

\smallskip
Radiative cooling is modeled according to rates from \citet{sutherland93} 
for gas with temperatures above $10^4\rm K$, and using rates from 
\citet{dalgarno72} for gas below $10^4\rm K$.
The metal-enriched gas is allowed to cool down to $0.1\rm K$. 
The gas is heated by a uniform UV background radiation following
\citet{haardt96}, which is turned on after reionization at $z\!=\!10$.
Optically thick regions with hydrogen number density greater than 
$n_{\rm H} \!=\! 0.01 \rm\,H\,cm^{-3}$ are assumed to be self shielded 
from the UV radiation, so the heating rate is reduced accordingly
by a factor of $\exp(-n/0.01  \rm\,H\,cm^{-3})$.

\smallskip
The star formation rate follows a Kennicutt-Schmidt relation 
\citep{kennicutt98},
$\rho_{\rm {SFR}} = \epsilon_{\rm ff} \rho_{\rm g}$,
where $\rho_{\rm {SFR}}$ is the star formation rate density, $\rho_{\rm g}$
is the gas density and $\epsilon_{\rm ff}$ is the varying star 
formation efficiency per free-fall time.
Stars form in cells with hydrogen number density above a threshold of
$n_0=10\rm\, H\,cm^{-3}$. 
The star formation efficiency per free-fall time ($\epsilon_{\rm ff}$) is
determined by the local turbulent Mach number and the Jeans length 
\citep{kimm17,trebitsch18}.
The minimal mass for a stellar particle is $10^4\rm\, M_\odot$, where
each such particle represents a population of stars with a Chabrier 
initial mass function \citep{chabrier05}, imposing lower and upper bounds 
of $0.1\rm\,M_\odot$ and $150\rm\, M_\odot$, respectively.

\smallskip
Stellar feedback is included in the simulation through Type-II supernovae,
each releasing kinetic energy of $10^{51}\rm\,erg$.
A stellar particle allows supernova explosions once older than $5\rm\,Myr$. 
The explosion deploys 31 per cent of the mass to the ISM 
with a metal yield of 5 per cent of the total ejecta.
Only massive stars above $6\rm M_\odot$ are assumed to explode. Although the
frequency of supernova explosions per solar mass is $0.015\rm\,M_\odot^{-1}$, 
this frequency is multiplied by a factor of two $0.03\rm\,M_\odot^{-1}$ to 
account for the effect of clustered supernovae, which may lower the density 
in the surrounding medium and increase the radial momentum per supernova in 
the subsequent explosions \citep{kim17,gentry17,gentry19}.
Mechanical supernova feedback is implemented \citep{kimm14,kimm15} to 
model separately the supernova energy-conservation phase and the 
momentum-conservation phase.
The model identifies the current phase of the supernovae explosion and 
directly deploys the expected momentum during the \emph{snow-plow} 
momentum-conserving phase.
In addition to the momentum from supernovae, the simulation takes into account 
a contribution to radial momentum due to the pre-heating in the vicinity of 
young OB stars \citep{geen15}.

\subsection{Black holes and AGN feedback}
\smallskip
Massive black holes are implemented as sink particles with a seed mass of 
$10^4\rm\, M_\odot$ and zero spin.
Sink particles are formed in cells where 
both the stellar and gas density exceed the density threshold for star 
formation, while the stellar velocity dispersion is larger than 
$20\rm\,km\,s^{-1}$, and the new seed is
located at least $50\rm\,ckpc$ away from a neighboring pre-existing 
black-hole.

\smallskip
The black-hole growth rate is assumed to be
$\dotMbh\!=\!(1\!-\!\epsr)\,\dot{M}_{\rm Bondi}$,
where $\epsr$ is the spin-dependent radiative efficiency that is computed in
the simulation on-the-fly and $\dot{M}_{\rm Bondi}$ is 
the Bondi-Hoyle gas accretion rate \citep{bondi44},
\begin{equation}
\dot{M}_{\rm Bondi} = 4\pi G^2  \frac{ M_{\rm bh}^2\bar{\rho} }
{(\bar{u}^2+\bar{c}_{\rm s}^2) ^{3/2}} \, ,
\end{equation}
where $\bar{\rho}$ is the averaged gas density, $\bar{u}$ is the average 
relative velocity of the gas with respect to the black-hole and 
$\bar{c}_{\rm s}$ is the average speed of sound.
These averaged gas properties in the vicinity of the black-hole are
measured within a radius of $4\Delta x$ ($\sim 150\pc$) \citep{dubois12}. 
The highest resolution is kept in the vicinity of
the black-hole at all times.
The accretion rate is limited by the Eddington limit,
\be
\dot{M}_{\rm Edd}=\frac{4\pi G m_{\rm p} M_{\rm bh}}{\epsr\sigma_{\rm T} c} \, ,
\ee
with $\sigma_{\rm T}$ the Thompson cross section.

\smallskip
An explicit drag force is exerted by the gas on the black-hole
in order to avoid oscillations of the sink particles due to resolution effects 
close to high density regions.
The sub-grid drag force model mimics the unresolved dynamical friction as 
exerted by the gas on the massive black-hole.
The drag force is measured from the gas quantities within the high 
resolution sphere of radius $4\Delta x$ around the black-hole.
The functional form of the drag force is:
\begin{equation}
F_{\rm DF} = 4\pi G^2 \alpha f_{\rm g}  
\frac{M_{\rm bh}^2 \bar{\rho}}{\bar{c}_{\rm s}^2} \, ,
\end{equation}
where $\alpha$ is the boost factor in high density regions, 
$\alpha\!=\!(n/n_0)^2$ when the density is high ($n\! >\! n_0$) 
and $\alpha=1$ otherwise.
The fudge factor $f_{\rm g}$ is a function of the Mach number allowing a more 
efficient drag force when the gas is supersonic with respect to the black 
hole \citep[following][]{ostriker99}. 

\smallskip
AGN feedback is modeled in two modes of high and low accretion rate.
The distinction between these two modes depends on the ratio of the
black-hole accretion rate and the Eddington rate 
$\chi=\dot{M}_{\rm bh}/\dot{M}_{\rm Edd}$ \citep{dubois12} 
with a transition at $\chi=0.01$.
The power released by the AGN is $\dot{E}_{\rm AGN}=\eta \dot{M}_{\rm bh} c^2$,
with a different form for the feedback efficiency $\eta$ in the two
feedback modes.
The \emph{"radio mode"} occurs when the black-hole accretion rate is
low, $\chi < 0.01$, during which AGN feedback is in the form of jets, namely, 
a bipolar release of mass, momentum and energy into a cylinder of radius 
$\Delta x$ and height $\pm \Delta x$ above and below the black-hole position.
%
Feedback efficiency in the radio-mode scales with the black-hole spin. 
This was modeled using a fourth-order polynomial fit to sampled results from
\citet{mckinney12} of magnetically chocked accretion discs. 
The black-hole spin tends to values larger than $\sim 0.7$ once it starts 
to accrete gas and gain mass. As a result, while the radio mode is active, the
high spin values translate to a typical feedback efficiency of 
$\eta\!\geq\! 0.6$ \citep[][fig.~3]{dubois20_nh}.
The jets are set in motion with a speed of
$10^4\rm\,km\,s^{-1}$ and their direction is aligned (anti-aligned) with the 
black-hole spin.
The \emph{"quasar mode"} is active during phases of high accretion rate,
$\chi \ge 0.01$, where the feedback deposits thermal energy within a sphere 
of radius $\Delta x$. The feedback efficiency in the quasar mode is coupled to 
the spin-dependent radiative efficiency 
($\epsilon_{\rm r}$), $\eta = 0.15\epsilon_{\rm r}$ \citep{dubois12}.

The black-hole spin, starting from zero, is followed on-the-fly in the
simulation, including the evolution of spin magnitude and direction during
both gas accretion and binary black-hole mergers.
The evolution of black-hole spin is of paramount importance in the models
implemented in the simulation.
The radiative efficiency of the AGN, $\epsilon_{\rm r}$, is spin dependent,
therefore, variations in the magnitude of the spin affect the Eddington
luminosity as well as the efficiency of both feedback modes. In addition,
variation in the direction of black-hole spin will determine the direction
of the jets during low accretion rate phases.
The radiative efficiency is defined as:
\be
\epsr = f_{\rm att}\, (1-e_{\rm {isco}})\, ,
\ee
where $e_{\rm {isco}}$ is the energy per unit rest mass of the
innermost stable circular orbit (ISCO),
which depends on the Schwarzschild radius and the spin of the black-hole.
The radiative efficiency is reduced while the black-hole is in the radio mode
by the factor $f_{\rm att}\!=\!\min(\chi/\chi_{\rm tran}, 1))$ following 
\citet{benson09}.
When the accretion is low, the model assumes that the energy deposited as jets 
is powered by the rotation of the black-hole \citep{blandford77}, therefore the
magnitude of the BH spin is decreasing when jets are active 
\citep[following][]{mckinney12}.
At high accretion rates, the magnitude of the spin evolves following 
\citet{bardeen70}, spinning up (down) when the accretion disc is aligned 
(anti-aligned).
The angular momentum of accreted gas will determine the evolution in the 
direction of the black-hole spin.
When an accretion disc is misaligned, torques due to Lense-Thirring effect
drive the inner parts of the disc to be aligned (or anti-aligned) with the 
black-hole spin, causing a warp in the disc. Eventually, the viscosity of the 
disc will force the disc and the black-hole spin to a new stable configuration 
of alignment (anti-alignment) with the total angular momentum as vectorial sum 
$\boldsymbol{J}_{\rm tot}=\boldsymbol{J}_{\rm bh} + \boldsymbol{J}_{\rm disc}$ 
\citep{king05}.
The unresolved disc is assumed to be a thin Shakura-Sunyaev accretion
disc \citep{shakura73}, which is aligned with the angular momentum of the 
feeding gas \citep[details in][]{dubois14_bh}.

\smallskip
Two black holes are merged when the distance between them is smaller than 
$4\Delta x$ and their relative velocity is lower than the escape velocity of 
the binary.
After a binary black-hole merger, 
the spin of the remnant black-hole is determined by 
the spins and orbital angular momentum of the two merging black holes 
\citep[analytic expressions in][]{rezzolla08}.

\subsection{Galaxy sample and measurements}

The identification of haloes is done using the AdaptaHOP halo finder 
\citep{aubert04} with the Most Massive Substructure Method \citep{tweed09}, 
where the density is smoothed over 20 particles, and
considering only haloes 
with more than 100 dark-matter particles.
The center of the halo is identified in the standard {NewHorizon}
analysis using a shrinking-sphere algorithm,
where the dark-matter center of mass is calculated at each iteration followed 
by a reduction of the radius by 10 per cent down to a minimum radius of 
$0.5\kpc$ \citep{power03}. The peak of the dark-matter density within this 
final volume is selected as the halo center.
The galaxy center is identified in a similar fashion using the stellar 
particles with the HOP halo finder, imposing a minimum number of 50 stellar
particles. 
Haloes in the zoom-in volume of {NewHorizon}
that are polluted by low-resolution dark-matter
particles from the embedding {Horizon-AGN} volume
are excluded from the sample.

\smallskip  
In order to achieve a higher accuracy for the galaxy center position,
necessary for the studies of the BH locking to the center,
we apply an additional re-centering procedure within the initial radius of 
$\sim\! 0.2\Rv$.
We find the center of mass recursively within a shrinking sphere, where the 
radius is reduced by 5 per cent at each iteration down to a minimum radius of 
$3.5\Delta x$ as long as the sphere contains more than 100 stellar particles.
Still, locating the center in low-mass galaxies remains a challenge since 
the center is ill-defined when the potential well is shallow and there is 
no distinct bulge.

\smallskip
We identify the central black-hole in each galaxy by selecting the most 
massive black-hole within a radius of $0.2 R_{\rm v}$ from
the galaxy center at the final redshift available for our current analysis,
$z \seq 0.73$. 
This is also the closest black-hole to the center in all the cases studied
here.
We then follow the black-hole back in time using its
unique ID. In the event of a binary black-hole merger, the post-merger
particle posses the identity of the more massive progenitor.

\smallskip
The bulge mass is measured by the total stellar mass in the spheroid using 
kinematic decomposition. 
Each star particle $i$ at radius $r_{i}$ from the center and with a 
speed $v_{i}$ is assigned a ratio: $j_{{\rm z}, i}/j_{{\rm c}, i}$,
where $j_{{\rm z}, i}$ is the specific angular momentum along the 
direction of the galaxy spin axis, and $j_{{\rm c}, i}=r_{i}v_{i}$ is 
the maximal specific angular momentum of the star. 
Bulge star particles are defined as those obeying the condition
$j_{{\rm z}, i}/j_{{\rm c}, i} < 0.7$ while the rest are considered as disc stars.

\smallskip
\def\sampleTable{
   \def\topruleTable{${\rm Galaxy \#}$  &  $\log\, M_{\rm s}$ & $\log\, M_{\rm gas}$   & $M_{\rm bh}$ }
   \def\sectopruleTable{$\ $ & $\rm [10^{10}\ M_\odot]$ & $\rm [10^{10}\ M_\odot]$  &  $\rm [10^{6}\ M_\odot]$}
   \def\thirdtopruleTable{$\ $ & $(z=2, 0.73)$ & $(z=2, 0.73)$ & $(z=2, 0.73)$ }
   \def\toprow{\topruleTable\\ \sectopruleTable\\ \thirdtopruleTable\\}
   \begin{table}
           \begin{center}
           \resizebox{\columnwidth}{!}{
           \begin{tabular}{ccccc}
           \toprow\hline
                    01 &       0.69 , 2.59 &        0.22 , 2.50 &        0.21 , 1.66   \\
                    02 &       6.88 , 10.63 &        2.49 , 1.11 &        0.15 , 28.39   \\
                    03 &       0.50 , 6.40 &        1.02 , 1.60 &        0.03 , 6.40   \\
                    06 &       2.08 , 5.32 &        0.64 , 1.46 &        0.30 , 1.77   \\
                    10 &       1.17 , 6.14 &        1.23 , 0.81 &        0.47 , 24.42   \\
                    18 &       2.63 , 4.28 &        0.99 , 0.19 &        0.15 , 6.44   \\
                    51 &       0.27 , 2.16 &        0.18 , 0.42 &        0.02 , 0.25   \\
                   142 &       0.26 , 1.02 &        0.34 , 0.29 &        0.03 , 0.09   \\
           \hline
           \end{tabular} 
           }
\caption[]{{Masses of stars, gas and black-hole
in the massive {NewHorizon} simulated galaxies at
$z\seq 2$ and $z\seq 0.73$}} 
\label{sample_table}
       \end{center}
   \end{table}
   }
\sampleTable
\section{Suppression by Supernovae}
\label{sec:app_Mbh_raw}

\begin{figure*} 
\centering
\includegraphics[width=1.0\textwidth]{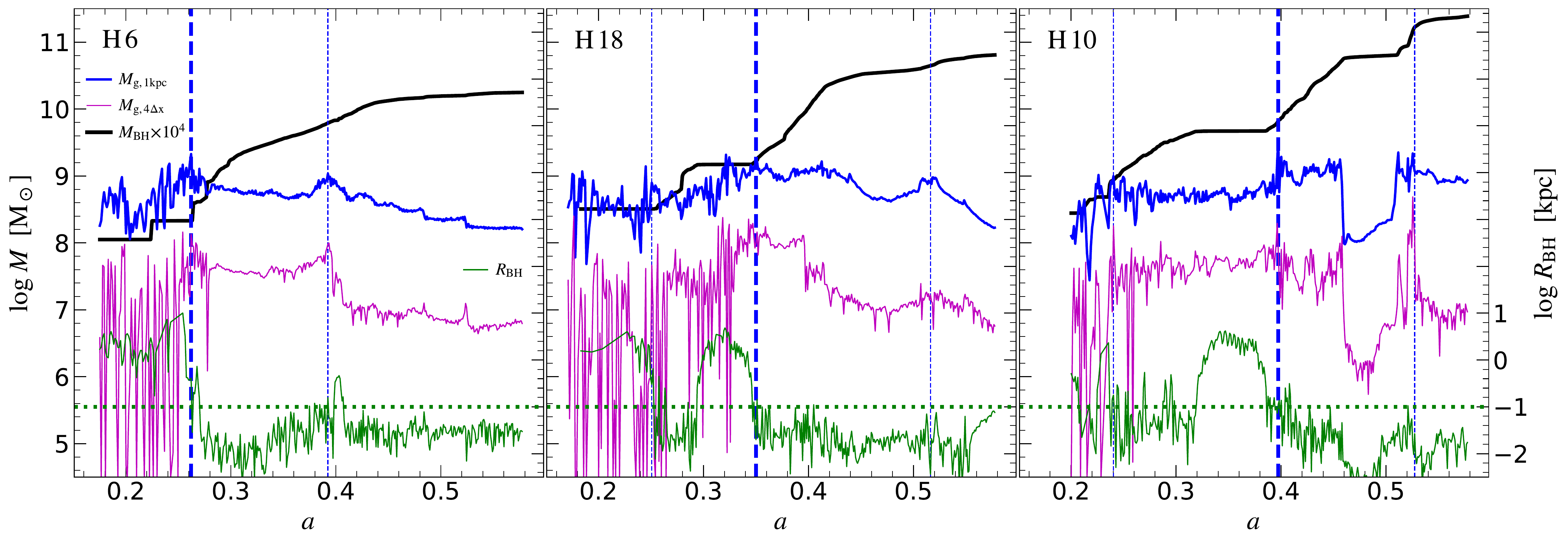}
\caption{
The evolution of three galaxies in the sample, as in \fig{Mbh}, 
but where the gas mass within $1\kpc$ (blue) is shown 
without smoothing every output timestep.
Also shown are 
the black-hole mass (scaled $\Mbh\!\times\! 10^{4}$, black), 
the position of the black-hole with respect to the center $R_{\rm BH}$ 
(green, right axis), and the gas mass within a smaller sphere of radius 
$4\Delta x$, namely four grid cells, or $\sim\! 150\pc$ about the galaxy center.
The strong fluctuations in the supernova zone before the major blue-nugget peak
(marked by the vertical thick dashed blue line) 
indicate multiple episodic supernova-driven gas ejections 
from the central $1\kpc$.
Galaxies H18 and H10, middle and right panel respectively, undergo a 
minor compaction event before the major blue-nugget peak 
(vertical thin dotted blue line).
}
\label{fig:Mbh_raw}
\end{figure*}

\Fig{Mbh_raw} displays the evolution of central gas mass ($M_{\rm g, 1\kpc}$) 
in three galaxies, as in \fig{Mbh}, but here showing the raw data at each output
timestep without any smoothing over time.
Also shown is the 
gas mass within a smaller sphere of radius $4\Delta x$ about
the galaxy center, four minimum grid cells or $\sim\! 150\pc$, within which
the gas properties are used to determine the accretion rate onto the black-hole
in the simulations.
Galaxy H6 in \Fig{Mbh_raw} (left panel) shows the most clear compaction event 
associated with an onset of rapid black-hole growth.
There are large central gas oscillations 
prior to the major blue-nugget peak ($a\ssim 0.26$). 
These fluctuations can be interpreted as due to repeating episodes of
supernova-driven gas ejection in the pre-compaction phase.
During the pre-BN phase we usually find the black-hole wandering away 
from the center with suppressed mass growth. 
Once the compaction has occurred and the potential well is deep enough to 
resist SN-driven gas expulsion from the center, we see a transition to 
smaller central gas mass fluctuations, the black-hole sinks to the inner 
$\sim\! 100\pc$ as can be seen in the green curve ($R_{\rm BH}$), 
and it is rapidly gaining mass.

\smallskip
In the middle panel of \fig{Mbh_raw} we show the evolution of 
galaxy H18 which generally displays a similar behaviour as seen in H6, 
namely fluctuations with large amplitudes before the major compaction peak 
at $a\ssim 0.35$, and smaller amplitudes in the post-BN phase.
Unlike H6, here we identify a minor compaction at $a\ssim 0.25$ 
before the major blue-nugget peak. 
Although the fluctuations remain relatively large after the minor compaction, 
for a short period of time, between $a\ssim 0.26-0.28$, 
we see a subtle indication for fluctuations with smaller amplitude, 
possibly caused by an abrupt gas accumulation in the nuclear region 
($4\Delta x$) which temporarily increase the depth of the potential well.
The gas supply to the center is diminished at $a\ssim 0.28$ 
as the compaction ends, and once again we see large fluctuations that persist 
until the major compaction.
This behaviour suggests that while the galaxy is still below the golden mass, 
a minor compaction may briefly induce a deeper potential well due to the 
sudden gas accumulation in the center, thus triggering the sinkage 
of the black-hole to the center as well as causing a temporary reduction 
in the gas fluctuations, both of which will assist the black-hole growth.
However, such episodes tend to be short lived when the galaxy is still in the 
regime where SN explosions can efficiently expel and heat much of the gas 
before it can form massive permanent stellar body in the center.

\smallskip
In galaxy H10, \Fig{Mbh_raw} right panel, we again identify a minor 
compaction event ($a\ssim 0.24$) before the major blue-nugget ($a\ssim 0.4$). 
Here the minor compaction is stronger compared to H18 
(or any other pre-BN minor compaction event in the sample), 
as is evident from the cyan line in \fig{Mbh} where we see that 
$V_1$ is reaching $\Vsn \ssim 120\kms$ shortly after the compaction, 
indicating that the potential well is close to the critical value 
below which the galaxy is in the regime of effective SN feedback.
We see here somewhat smaller gas fluctuations after the minor compaction.
Although this compaction is relatively strong, by the end of this 
event at $a\ssim 0.28$, the potential well is not deep enough to lock the 
black-hole to the gas-rich center, and after $a\ssim 0.32$ it starts 
wandering off center until the major compaction event.

\smallskip
In all cases we find a correlation between the 
sinkage of the black-hole to the center and a compaction event, 
be it the major compaction or a minor one. 
However, it is only after the major compaction, when the galaxy has reached the 
golden mass, that we find a potential well which is deep enough to both lock the
black-hole to the galactic center and permanently reduce gas fluctuations.

We do not find a prolonged period of time in which the black-hole is in the 
center while the fluctuations are large. This is not surprising if both the 
black-hole sinkage to the center and the reduction in gas fluctuations have 
the same origin, namely the deepening of the potential well due to a 
compaction event.
As a result, we cannot distinguish here the possible effect of 
gas fluctuations on the suppression of black-hole growth 
from suppression caused by off centered wandering. 
The effect of central gas fluctuations on suppression of black-hole growth 
will require further investigation beyond the scope of this paper.

\smallskip
A similar transition from large to small oscillations in the central gas
density within $\sim\!100\pc$ has been seen in FIRE simulations,
Fig.~1 of \citet{angles17}. The large fluctuations are seen at early times
when the black-hole growth is suppressed, while the black-hole starts to grow
when the central surface density becomes stable.
The repeated ejections in the pre-compaction supernova regime may indicate
that certain suppression of black-hole growth could have occurred even if the
black-hole was at the center at that time and not wandering off center.
This implies that the suppression of black-hole growth 
may not be solely due to its wandering off center.
%
On the other hand, the intense episodes of supernova feedback may themselves be
responsible for the wandering of the black-hole off center. 
The continuous gas removal from the center may keep the central 
potential well shallow, 
allowing the black-hole to orbit about the center. This is until the potential
well is deepened by the compaction event, which forces the black-hole to
sink to the center, where it can accrete more efficiently.

\smallskip
Although off-centered black holes have been observed in dwarf galaxies
\citep{Reines20}, 
we caution that further study is required to 
examine the validity and causes for wandering black holes in simulations.

\section{More Images}
\label{sec:app_images}

\begin{figure*} 
\centering
\includegraphics[width=0.874\textwidth,trim={0cm 1.69cm 0cm 0cm}]
{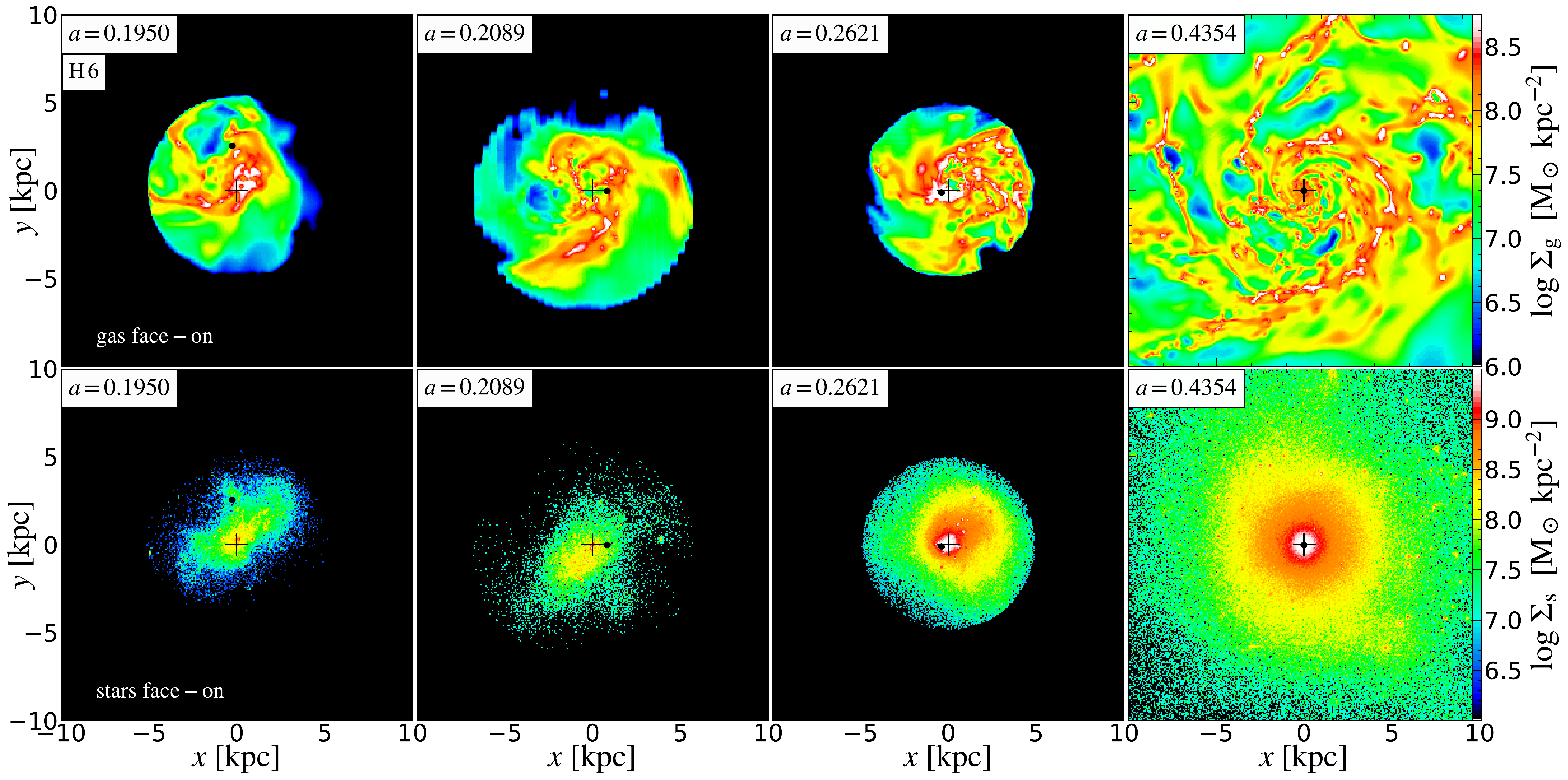}
\includegraphics[width=0.874\textwidth,trim={0cm 1.69cm 0cm 0cm}]
{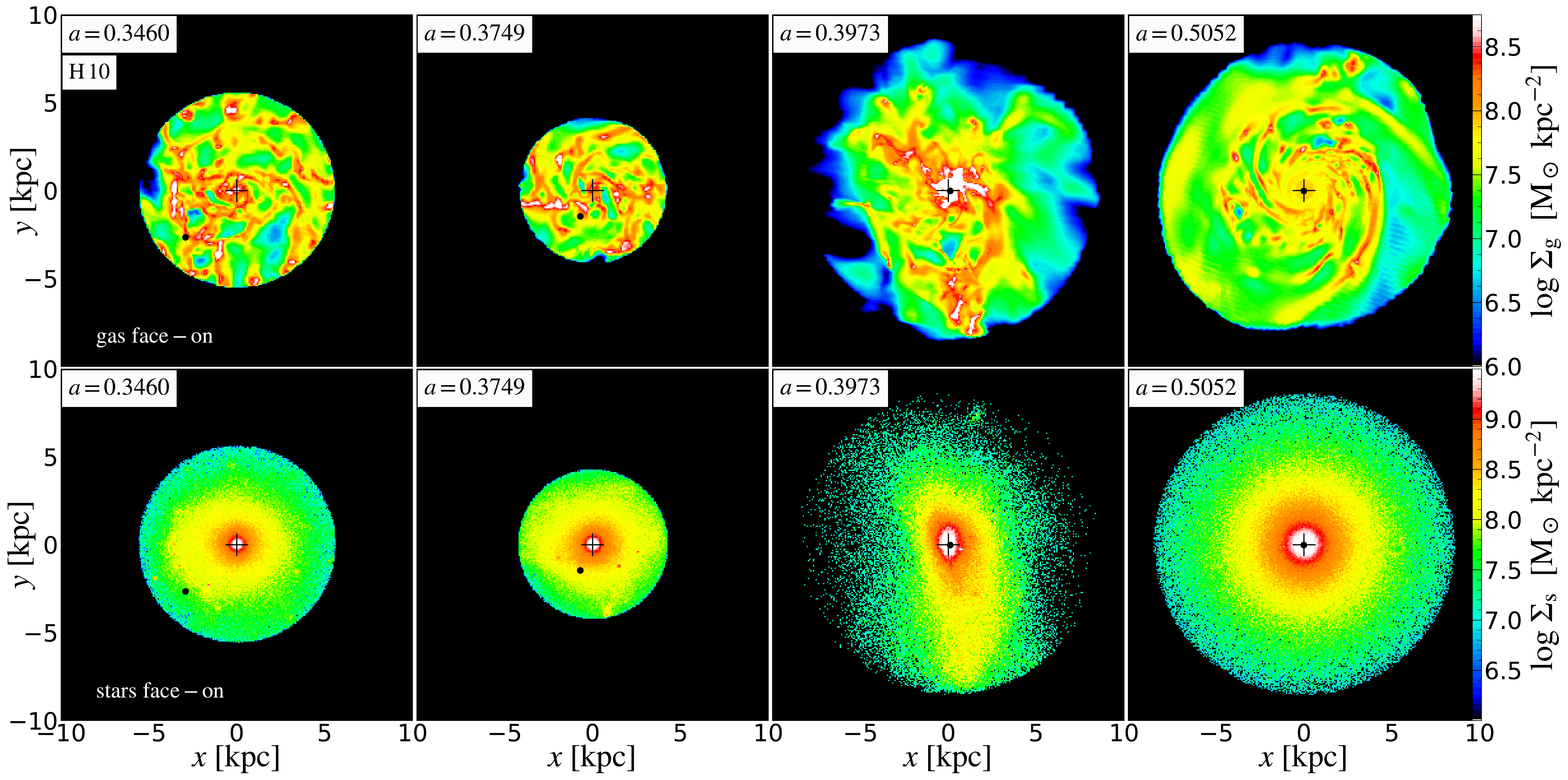}
\includegraphics[width=0.874\textwidth]
{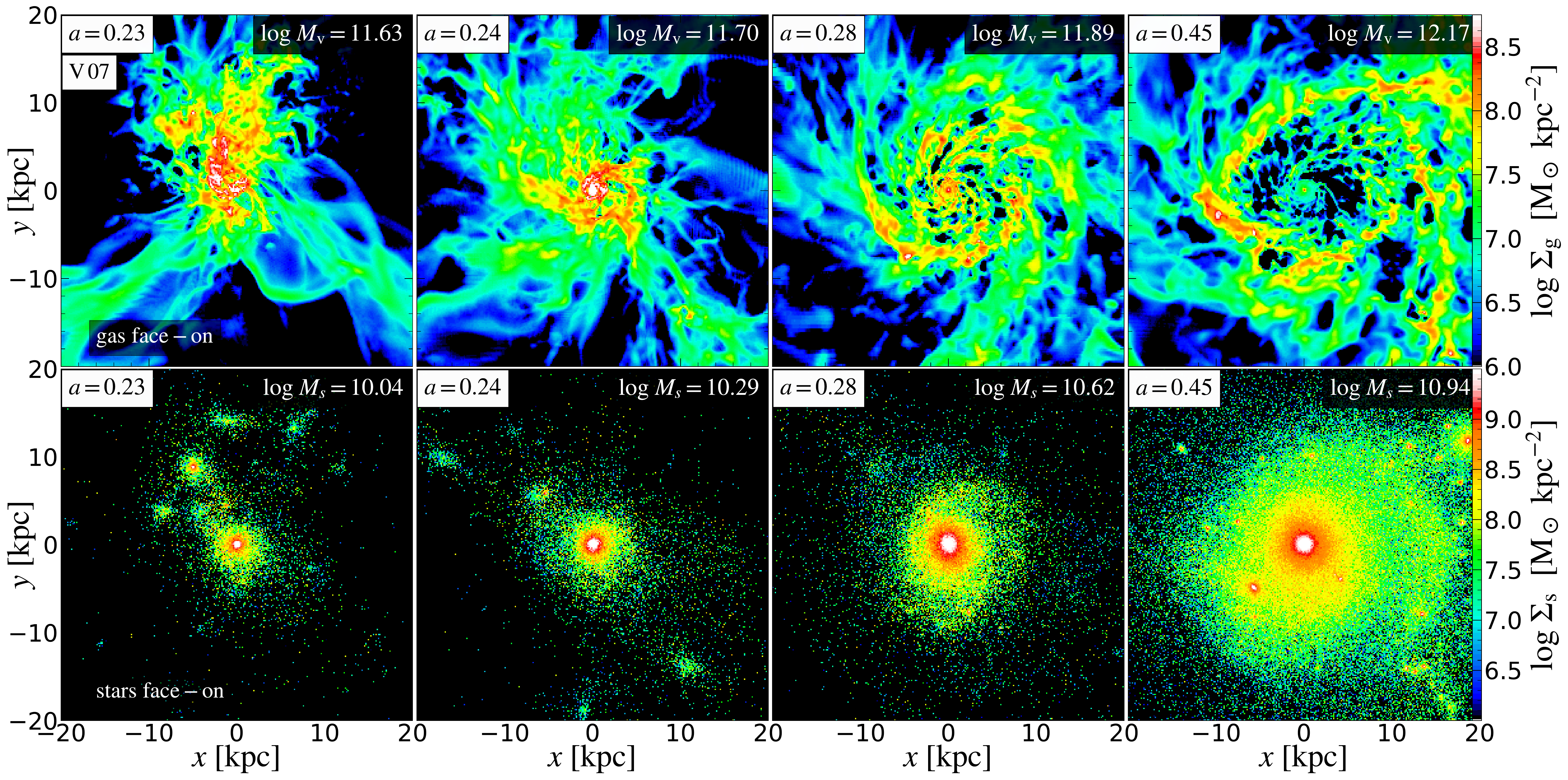}
\caption{
{\bf Top \& Middle:} Similar to \fig{mosaic_H2} but for {NewHorizon} 
galaxies H6 and H10 respectively.
{\bf Bottom:} Similar to \fig{mosaic_H2} but for {VELA} galaxy V07.
From left to right.
First: during the compaction process
($\log \Ms \seq 10.0$, $\log \Mv\seq 11.6$).
Second: at the blue-nugget phase (10.3, 11.7).
Third: post-compaction VDI disc (10.5, 11.8).
Forth: post-compaction, clumpy, long-lived ring,
fed by incoming streams (10.8, 12.1).
}
\label{fig:mosaic_H6_H10_V07}
\end{figure*}

\Fig{mosaic_H6_H10_V07} top and middle panel
present a sequence of images similar to \fig{mosaic_H2} 
for two additional {NewHorizon} galaxies, H6 and H10 respectively.   
\Fig{mosaic_H6_H10_V07} bottom panel brings an analogous set of images 
for a {VELA} galaxy, V07 \citep[e.g.][]{zolotov15}.
The sequence of images, for face-on gas and stars, demonstrate the
pre-compaction stage, the compaction process, the peak gas compaction in a
blue-nugget phase, and the post-compaction disc and ring about a compact
passive red nugget.
The evolution is qualitatively similar in {VELA} and New Horizon, despite the
different codes and different ways by which the physical sub-grid recipes for
star formation and feedback are incorporated.
The {NewHorizon} images show the position of the black-hole in the different
stages, demonstrating that it tends to wander off center before the compaction
and sink to the center during the compaction process.

\bsp	
\label{lastpage}
\end{document}